\documentclass[
prx,
superscriptaddress,
twocolumn,
floatfix,
longbibliography
]{revtex4-1}

\usepackage{float}
\usepackage{bbm}
\usepackage{epsfig}
\usepackage{epstopdf}
\usepackage{graphicx}
\usepackage[caption=false]{subfig}
\usepackage{amssymb}
\usepackage{amsmath}
\usepackage{scalefnt}
\usepackage{color}
\usepackage[title]{appendix}
\usepackage[pdftex,letterpaper=true,pagebackref=false]{hyperref}
\usepackage{qcircuit}
\usepackage[capitalize]{cleveref}
\usepackage{braket,mleftright}
\usepackage{physics}
\usepackage{import}
\usepackage{dcolumn}
\usepackage{csquotes}
\usepackage{soul}

\usepackage[usenames,dvipsnames]{xcolor}
 
\floatstyle{plaintop}
\restylefloat{table}

\hypersetup{
pdftitle={Measurement Error Mitigation for Variational Quantum Algorithms},
pdfauthor={George S. Barron and Christopher J. Wood},
	pdfnewwindow=true, 
	colorlinks=true, 
	linkcolor=MidnightBlue,
	citecolor=Magenta,
	urlcolor=RoyalBlue
}

\begin{document}

\title{Measurement Error Mitigation for Variational Quantum Algorithms}

\author{George S. Barron}
\email{gbarron@vt.edu}
\affiliation{Department of Physics, Virginia Tech, Blacksburg, VA 24061, U.S.A}

\author{Christopher J. Wood}
\email{cjwood@us.ibm.com}
\affiliation{IBM Quantum, IBM T.J. Watson Research Center, Yorktown Heights, NY 10598, USA}

\date{\today}

\begin{abstract}
    Variational Quantum Algorithms (VQAs) are a promising application for near-term quantum processors, however the quality of their results is greatly limited by noise.
    For this reason, various error mitigation techniques have emerged to deal with noise that can be applied to these algorithms. Recent work introduced a technique for mitigating expectation values against correlated measurement errors that can be applied to measurements of 10s of qubits. We apply these techniques to VQAs and demonstrate its effectiveness in improving estimates to the cost function. Moreover, we use the data resulting from this technique to experimentally characterize measurement errors in terms of the device connectivity on devices of up to 20 qubits. These results should be useful for better understanding the near-term potential of VQAs as well as understanding the correlations in measurement errors on large, near-term devices.
\end{abstract}

\maketitle

\section{Introduction}\label{sec:intro}
  The development of quantum computers and their applications have rapidly accelerated over the last few years. Several different hardware platforms have been experimentally realized at varying scales \cite{cross2019validating, pino2020demonstration, karalekas2020quantum, jurcevic2020demonstration}, and there has been an increased focus on studying algorithms and applications that can be run on these noisy near-term devices.
  Some of the most promising algorithms for near-term devices are Variational Quantum Algorithms (VQAs)~\cite{farhi2014quantum, Peruzzo2014, OMalley2016, McClean2016, kandala2017hardware, Colless2018, pagano2019quantum, romero2019variational},
  which use a quantum device to evaluate an objective function that is minimized using a classical optimizer. Instances of this algorithm include the Variational Quantum Eigensolver (VQE)~\cite{Peruzzo2014, OMalley2016, McClean2016, kandala2017hardware, Colless2018, ollitrault2019quantum, arute2020hartree}
  and Quantum Approximate Optimization Algorithm (QAOA)~\cite{farhi2014quantum, otterbach2017unsupervised, hadfield2019from}.
  Recent process in this field includes, for example, improvements to the measurement process~\cite{babbush2018low, verteletskyi2019measurement, huggins2019Efficient, zhao2020measurement},
  selection of variational ans\"atze ~\cite{Grimsley2018, Barkoutsos2018, tang2019qubitadaptvqe, gard2019efficient}, and optimization. Moreover, they have been demonstrated experimentally on a variety of physical systems~\cite{kandala2017hardware, Zhueaaw9918, dumitrescu2018cloud, klco2018quantum}.

  The ability of current experimental implementaions of VQAs to produce accurate results is limited by noise on the device, despite these algorithms not explicitly requiring error correction. All proposed platforms for quantum computation experience some combination of different errors including decoherence, calibration errors, leakage, cross-talk, and measurement errors. With superconducting systems, cross-talk and measurement are among the largest sources of errors \cite{chen2019detector}.
  Error mitigation techniques have been developed for VQAs to reduce the amount of error on near-term devices in the absence of error correction. For example, \emph{extrapolation to the zero-noise limit}~\cite{temme2017error,kandala2019error} uses pulse-level control to mitigate expectation values against decoherence, requiring only a constant factor of overhead in the number of circuits executed. Techniques have also been developed to characterize and mitigate against measurement errors
  \cite{hamilton2020error, geller2020efficient, hamilton2020scalable}. Moreover, Ref.~\cite{geller2020rigorous} analyzes the underlying model and provides rigorous improvements to correction techniques. Related techniques have also been used in VQE experiments \cite{kandala2017hardware} and are implemented in the IBM Qiskit package \cite{Qiskit}.
  Recently, Ref.~\cite{bravyi} has introduced a readout error mitigation technique, which we will call Continuous-Time Markov-Process Error Mitigation (CTMP-EM), that mitigates expectation values against correlated measurement errors.
  The $n$-qubit calibration procedure for CTMP-EM requires as few as $n+2$ circuits to execute and $O(n^2)$ parameters to fit.
  In this work, they used it to mitigate estimates of the fidelity of graph states using stabilizer measurements, as well as expectation values of stabilizers with respect to Clifford circuits.

  In this paper we apply the CTMP-EM technique to experimentally characterize long-range correlations in measurement errors, and to calibrate error mitigated measurements in several quantum computers.
  In addition to demonstrating the presence of long-range correlations in these devices, we are also able to show that these long-range correlations can be as strong between distant qubits as they are between neighboring qubits. Moreover, rather than only considering the global minimum for the VQE objective function, we consider the objective function holistically. Evaluating the objective function at other parameter values is important for various tasks, for example the application of the ubiquitous parameter shift rule \cite{PhysRevA.99.032331} for analytic gradient computation in variational experiments. For the Fermi-Hubbard model, we demonstrate that CTMP-EM can fundamentally improve the shape of the objective function for the VQE not only at its minimum, but globally as well.

  This article is structured as follows. In \cref{sec:background} we review VQAs and the CTMP-EM technique. In \cref{sec:mitvqa} we demonstrate that applying CTMP-EM to the VQE algorithm changes the shape of the objective function. In \cref{sec:char} we use the calibration data from CTMP-EM to analyze the long-range correlations in readout errors on devices. We also compare several different IBM Quantum superconducting devices with the CTMP-EM calibration data. In \cref{sec:conclusion} we conclude.

\section{Background}\label{sec:background}
  \subsection{Variational Quantum Algorithms}\label{sec:vqa}

    A VQA is an optimization problem $\min_\theta f(\theta)$ where the objective function $f(\theta)$ is evaluated using a quantum device within the classical optimization loop.
    Two examples of VQAs are the Variational Quantum Eigensolver (VQE) and the Quantum Approximate Optimization Algorithm (QAOA). In each case, the objective function is $f(\theta) = \bra{\psi(\theta)} H \ket{\psi(\theta)}$ for a Hamiltonian $H$ and parameterized state $\ket{\psi(\theta)}$. In the case of a VQE, $H$ corresponds to the Hamiltonian of some physical system. In the case of QAOA, $H$ corresponds to a cost function with binary variables. In QAOA, the state $\ket{\psi(\theta)}$ is prepared by alternating unitary gates corresponding to evolution of the cost Hamiltonian $H$ and some mixer operator(s). In VQE, the state $\ket{\psi(\theta)}$ takes the form of a specific trial-state ansatz.
    The objective function $f(\theta)$ is computed by performing measurements to estimate the expectation value $\bra{\psi(\theta)} H \ket{\psi(\theta)}$ on a quantum device.

    VQAs are appealing for near-term devices as they are agnostic to errors in state preparation so long as the true minimum expecation value can be reached. This has been demonstrated on different hardware platforms, with different instances of VQAs, and varying degrees of accuracy.
    For the case of VQE it has been shown that, in special cases, the parameters that minimize the objective function are resilient to certain types of noise \cite{sharma2019noise}.
    In general, however, device noise significantly impacts the value of the objective function at that minimum and other points in parameter space.

  \subsection{Continuous Time Markov Process Error Mitigation}\label{sec:ctmp}
    Error mitigation techniques generally aim to improve the accuracy of the results obtained from using a \emph{noisy} quantum device. Typically, each technique mitigates against a certain kind of error.
    Measurement errors are a large source of error in VQA experiments on near-term devices.
    Measurement errors are modeled by a stochastic \emph{assignment matrix} $A$ acting on the state before readout. The elements of the matrix $A_{y,x} = P(y|x)$ are the probability of reading out the basis state $y$ where $x$ was prepared.
    A general $n$-qubit stocastic matrix has $2^n(2^n-1)$ independent real parameters and thus can only be completely characterized for a small number of qubits.
    Once this matrix is determined, applying the inverse matrix to a given probability distribution undoes the effect of readout error, however since the inverse is not in general a stochastic matrix the resulting output is not a valid probability distribution.

    The CTMP-EM algorithm of Ref.~\cite{bravyi} circumvents dealing with the $A$ matrix directly, and is used to mitigate expectation values computed with a quantum device.
    This is done by modeling $A = e^G$, where $G = \sum_{i} r_i G_i$ with rates $0 \leq r_i \in \mathbb{R}$ and operators $G_i$ that generate different readout errors. In particular, for multi-qubit bitstrings $a$, $b$, the readout error $a \rightarrow b$ corresponds to the generator $G_i = \ket{b}\bra{a} - \ket{a}\bra{a}$. For readout errors $0 \leftrightarrow 1$, $01 \leftrightarrow 10$, and $11 \leftrightarrow 00$ on subsets of $n$-qubits, CTMP-EM can determine the corresponding $r_i$ with as few as $n+2$ circuits.
    Once $r_i$ have been determined, Algorithm 1 of Ref. \cite{bravyi} can be used to estimate measurement error mitigated expectation values. The idea of this algorithm is to use the measurement counts collected from an expectation value experiment and classically post-process them by simulating a Markov process that applies $A^{-1}$ to the resulting distribution while simultaneously computing the expectation value of the desired operator. This algorithm runs in $n^2 \gamma e^{4 \gamma} / \delta^2$ time, where $n$ is the number of qubits, $\delta$ is the desired additive error in the expectation value of the Pauli string considered, and
    $\gamma = - \text{max}_x \bra{x} G \ket{x}$ for bitstrings $x$. Ref.~\cite{bravyi} shows experimentally that $\gamma \approx 0.05 n$ for up to $n=20$. Hence, the algorithm requires exponential post-processing time, but in practical situations this technique is applicable up to about 50 qubits.

    To study the effects of measurement error mitigation on VQAs we perform simulations of noisy measurements which simulate the device including only readout errors present on individual qubits and all qubit pairs. We choose readout error rates to reflect the \emph{ibmq\_boeblingen} device. However for device characterization in \cref{sec:char} we use the real IBM Quantum devices.

\section{Mitigating VQA Objective Functions}\label{sec:mitvqa}

  \subsection{Ground State Energy Mitigation}\label{sec:gsemit}

    One of the main applications of VQE is to estimate the ground state energy of a Hamiltonian by minimizing the measured operator expectation value over the variational parameters.
    Both the expectation value and gradient estimates are particularly senstive to measurement errors which can greatly effect the the performance of the classical optimizer that depends on these values. This makes measurement error mitigation essential for improving the accuracy of VQE and other VQAs.
    To investigate the impact of CTMP-EM on variational algorithms, we choose the Fermi-Hubbard model 
    which describes a system of Fermions interacting on a lattice~\cite{hubbard1963electron}. The Hamiltonian is
    \begin{equation}
	     H = - t \sum_{\left< j, k \right>} \sum_\sigma \left( a^\dagger_{j,\sigma} a_{k,\sigma} + a^\dagger_{k, \sigma} a_{j, \sigma} \right) +
	      U \sum_{k} n_{k,\uparrow} n_{k, \downarrow},
    \end{equation}
    where $t$ is the tunneling parameter, and $U$ is the interaction parameter between Fermions on the same site,  $a^\dagger_{k,\sigma}$ is the raising operator for site $k$ with spin $\sigma \in \left\{\uparrow, \downarrow\right\}$, and $n_{k,\sigma} = a^\dagger_{k,\sigma} a_{k,\sigma}$ is the number operator. The Fermionic operators are mapped to qubit operators using the Bravyi-Kitaev mapping~\cite{BRAVYI2002210}. For all calculations we will assume $U = 2t$. Energies are expressed in units of $t$. We assume that the lattice is a 1-D chain coupled by nearest neighbors with periodic boundary conditions.
    We chose an $n$-qubit variational ansatz $\ket{\psi(\theta)}$ consisting of an initial state $\ket{+}^{\otimes n}$ followed by six repititions of a layer of parameterized single-qubit $Y$-rotations followed by a layer of CZ gates between neighbouring qubits.
    
    \begin{figure}
    \centering
    \includegraphics[width=\columnwidth]{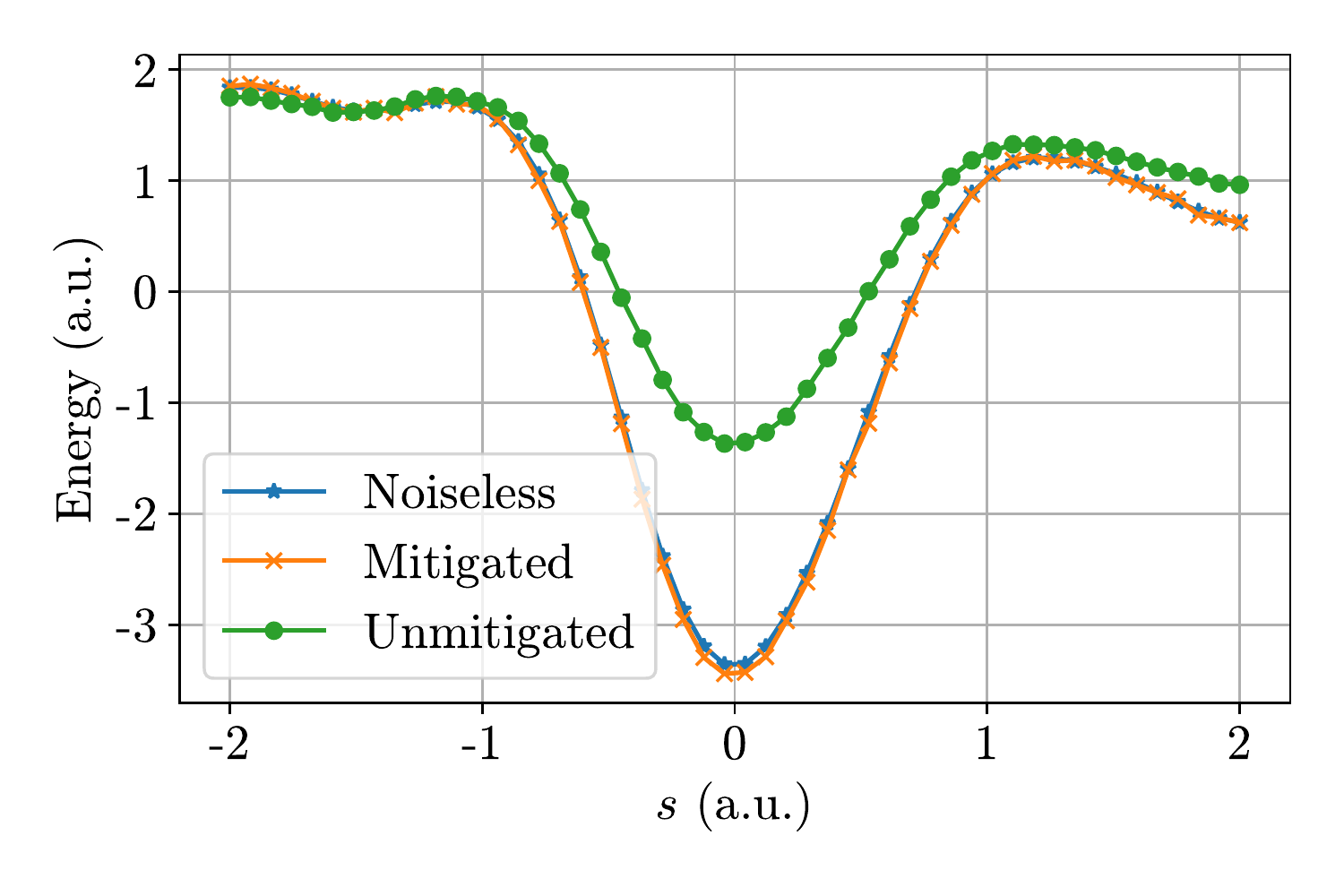}
    \caption{Sweep of the objective function $f$ through the global minimum ($s=0$) computed several ways. Here, $f$ corresponds to the energy of the Fermi-Hubbard model for the given state. In the ``Noiseless" case, there is no error in the simulation and no mitigation. In the ``Unmitigated" case, we include readout error, but no mitigation. In the ``Mitigated" case, we include readout error and CTMP-EM. Estimates of the ground state energy ($s=0$) and surrounding points are significantly improved by applying CTMP-EM.}
    \label{fig:sweep}
    \end{figure}

    In \cref{fig:sweep} we plot the objective function $f(\theta_0 + s \ \phi)$ around the global minimum, where $s$ is the parameter of the sweep, $\phi$ is a randomly chosen vector with $\dim(\theta_0) = \dim(\phi)$, and $\theta_0$ are the parameters that globally minimize $f$.
    Evaluations of $f$ are repeated three times: in the absence of any noise (``Noiseless"), including measurement error without any mitigation (``Unmitigated"), and including measurement error with CTMP-EM applied (``Mitigated").
    We observe that CTMP-EM is able to significantly improve both the estimate of the ground state energy, as well as objective function values around the ground state.
    Nevertheless, arriving at the correct ground state depends not only on the point containing the ground state itself, but the objective function as a whole.
    \begin{figure}
    \centering
    \subfloat{\includegraphics[width=\columnwidth]{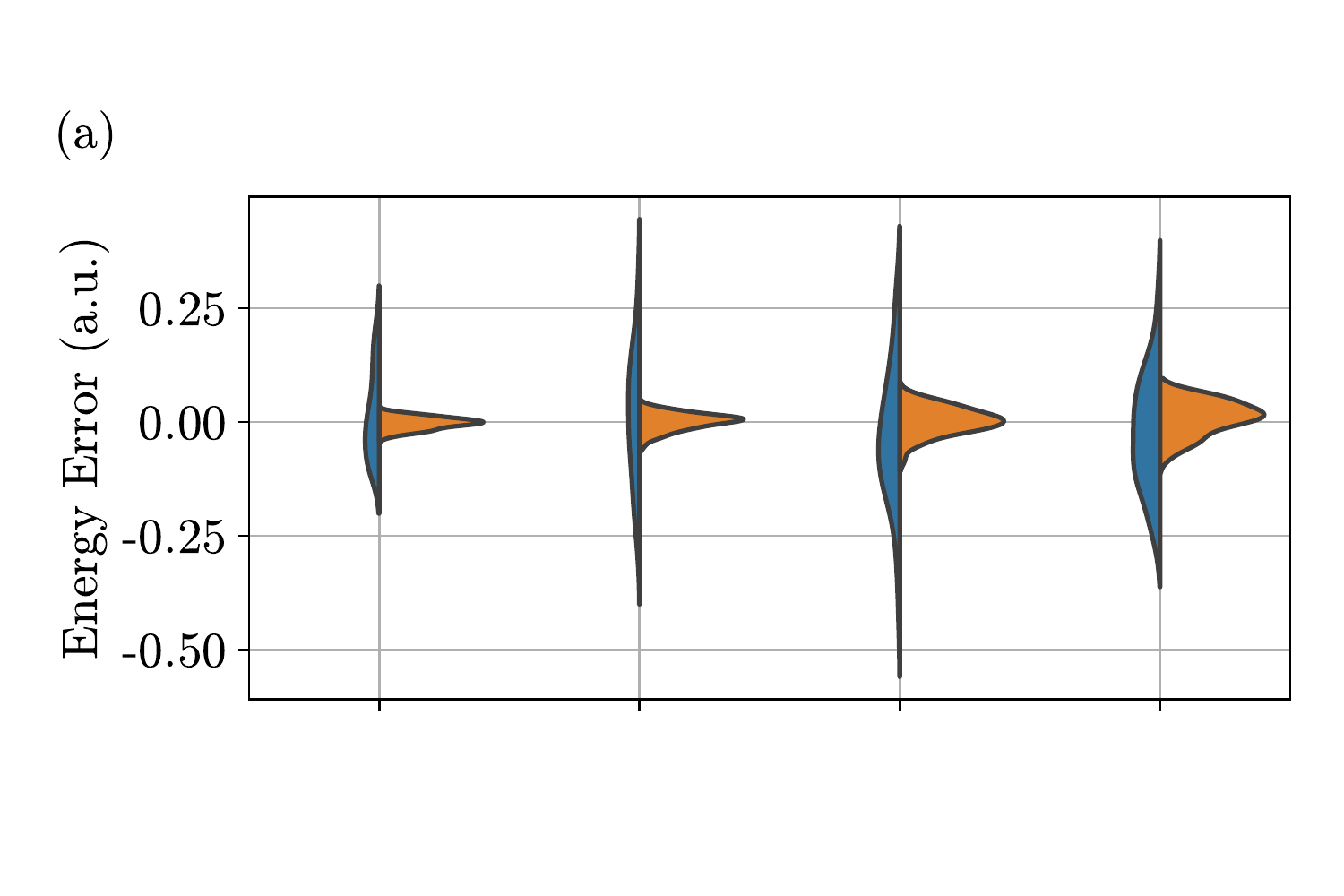}}\vspace{-1.0in}\\
    \subfloat{\includegraphics[width=\columnwidth]{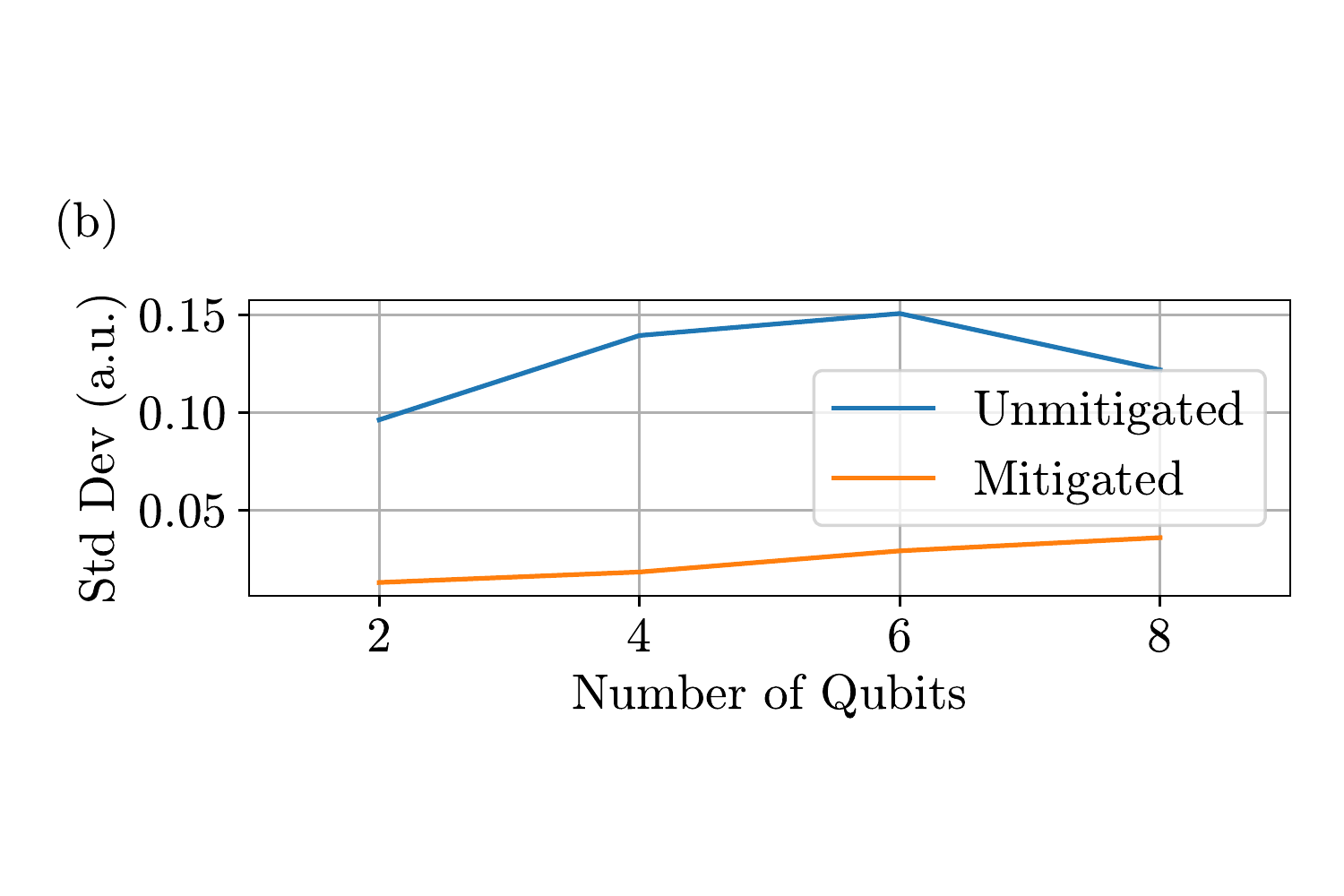}}
    \caption{Comparison of samples of the objective function $f(\theta)$ for random $\theta$ and different numbers of qubits. The energy error in panel (a) is the difference between the noisy (mitigated or unmitigated) energy and the exact result. The lefthand distributions are using the unmitigated objective function, and the righthand distributions are using the objective function mitigated with CTMP-EM. The standard deviations in panel (b) are those of the distributions in panel (a). For all system sizes considered, adding CTMP-EM significantly improves the estimate of the objective function. We use $8192n$ shots (for $n$ qubits) to compensate for the overhead in applying CTMP-EM. The remaining source of error in the mitigated case is due to undersampling in the number of shots required to perform CTMP-EM. This emphasizes the importance of the scaling of measurements needed for CTMP-EM with the number of qubits.}
    \label{fig:samp}
    \end{figure}
  
  \subsection{Objective Function Sampling}\label{sec:sampling}
    In VQE the quantum computer is treated as a black box evaluation of the objective function $f$, hence it is critical that evaluations of $f$ are accurate.
    To investigate the effectivness of measurement error mitigation for black box evaluations of $f(\theta)$ we sample points in parameter space $\theta$ and compare the noiseless, unmitigated, and mitigated cases for Fermi-Hubbard models with 1, 2, 3, and 4 sites (with 2, 4, 6, and 8 qubits respectively). We evaluate the energy with and without CTMP-EM for randomly sampled parameters of the objective function and compare the distribution of values with the noiseless result as shown in \cref{fig:samp}.
    \begin{figure*}
    \includegraphics[width=\textwidth]{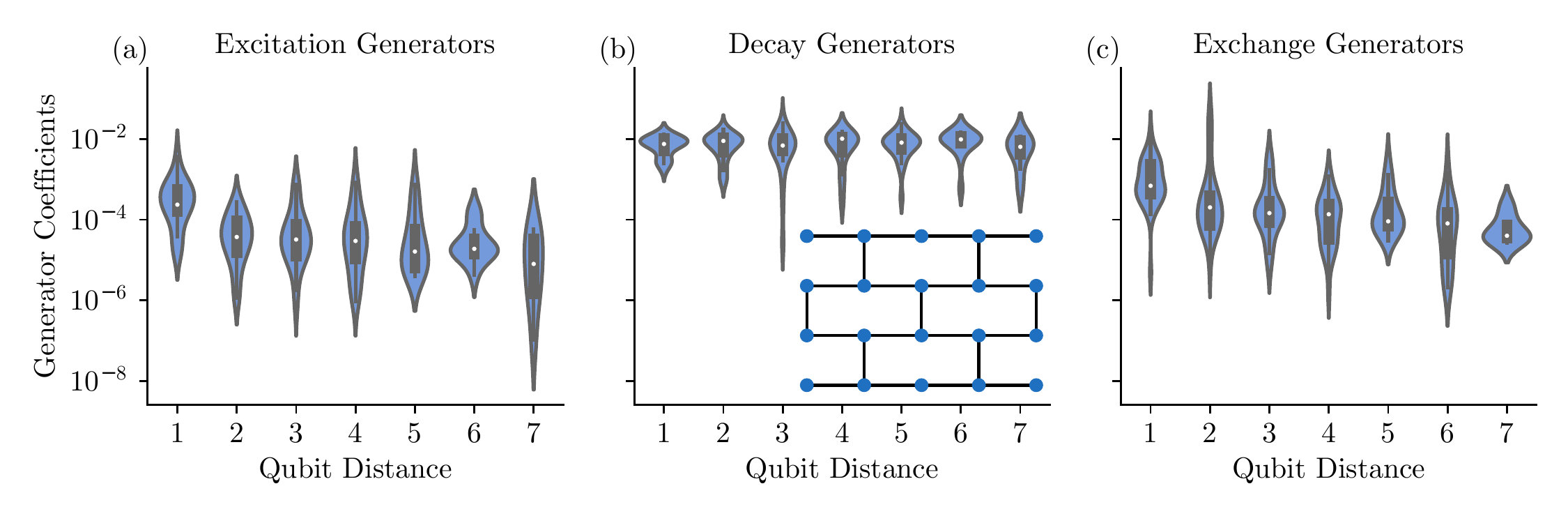}
    \caption{Histograms and quartiles of coefficients $r_i$ for various generators $G_i$. The ``Excitation" (a), ``Decay" (b), and ``Exchange" (c) generators are associated with readout errors $00 \rightarrow 11$, $11 \rightarrow 00$, and $01 \rightarrow 10$, respectively. The ``Qubit Distance" is the length of the shortest path between two qubits on \emph{ibmq\_boeblingen} device. Higher generator coefficients correspond to higher readout error rates. In all cases, the long-range correlations in readout error can be non-trivial. Cases where $r_i = 0$ due to shot limitations are elided for the purposes of plotting with logarithmic axes. In panel (b) we include the connectivity for the \emph{ibmq\_boeblingen} 20 qubit device. The vertices represent qubits, and the edges represent connections between qubits. Despite qubits being physically separated on the device, long-range correlations between qubits can occur.}
    \label{fig:char}
    \end{figure*}
  
    We find that the standard deviation of the error distribution is significantly reduced when mitigation is applied. Specifically, for 2, 4, 6, and 8 qubits respectively, the standard deviation is reduced by factors of approximately $7.46$, $7.64$, $5.18$, and $3.40$.
    This demonstrates that in the absence of CTMP-EM, the noisy objective function deviates significantly from its noiseless form which can greatly limit the effectivness of the VQE algorithm even for a small number of qubits.

\section{Characterizing Readout Errors}\label{sec:char}
    The CTMP-EM method can also be used as a characterization protocol for correlated measurement errors in a quantum device, which we will demonstrate by using it to characterize correlated readout errors in several experimental devices.
    One method for measurement error characterization involves computing the full $A$ matrix (which has $2^{2n}$ elements) as a form of measurement tomography \cite{bialczak2010quantum}, however this is not possible to do past a small number of qubits. Instead we use the set of rates $\{r_i\}$ of the CTMP-EM generator $G$ to characterize correlated measurement errors.
    This is scalable in the sense that it requires as few as $n+2$ circuits, and $G$ is parameterized by the $O(n^2)$ rates of its generator components.

  \subsection{Characterizing Correlated Errors}\label{sec:charcor}
    Calibration of $G$ in the CTMP-EM model described in Ref.~\cite{bravyi} is done by preparing a input set of computational basis states $\{\ket{a_i}\}$, labelled by bitstrings $a_i$, and performing measurements in the computational basis to estimate the assignment probabilities $P(x|a_i)$ for $x=0,...,2^n-1$. These probabilities are then processed to compute the CTMP-EM generator rates $\{r_i\}$ for each of the 1 and 2-qubit generator terms. The set of input labels $a_i$ is \emph{complete} if the set of all measurement outcomes contains all 1 and 2-qubit transitions for the CTMP-EM generators, if only 2-qubit correlations are present. This requires at least $n+2$ generators, though more may be used to provide a more uniform distribution across generator terms. We use the set of all bitstrings that have Hamming weight $\leq 2$, of which there are $(n^2 + n + 2) / 2$.
    For the purposes of characterizing correlated errors, we focus on the 2 qubit generators, as these generate correlations that cannot be captured in the single-qubit tensor product error model.
    We compare the distributions of the 2-qubit of rates $r_i$ grouped by the \emph{qubit distance}, which we define as the shortest path between two qubits in terms of the device connectivity.

    We apply this technique to the 20 qubit \emph{ibmq\_boeblingen} device, which has a planar qubit connectivity graph as shown inset in \cref{fig:char}. Here, for example, neighboring qubits have distance 1, and some pairs of qubits in the corners of the layout have distance 7.
    The histograms of measured 2-qubit generator rates $r_i$ vs qubit distance for the \emph{ibmq\_boeblingen} device are shown inset in \cref{fig:char}. Here we further group the generators into three types: excitation generators ($00 \rightarrow 11$), decay generators ($11 \rightarrow 00$), and exchange generators ($01 \leftrightarrow 10$).
    
    \begin{figure}
    \includegraphics[width=\columnwidth]{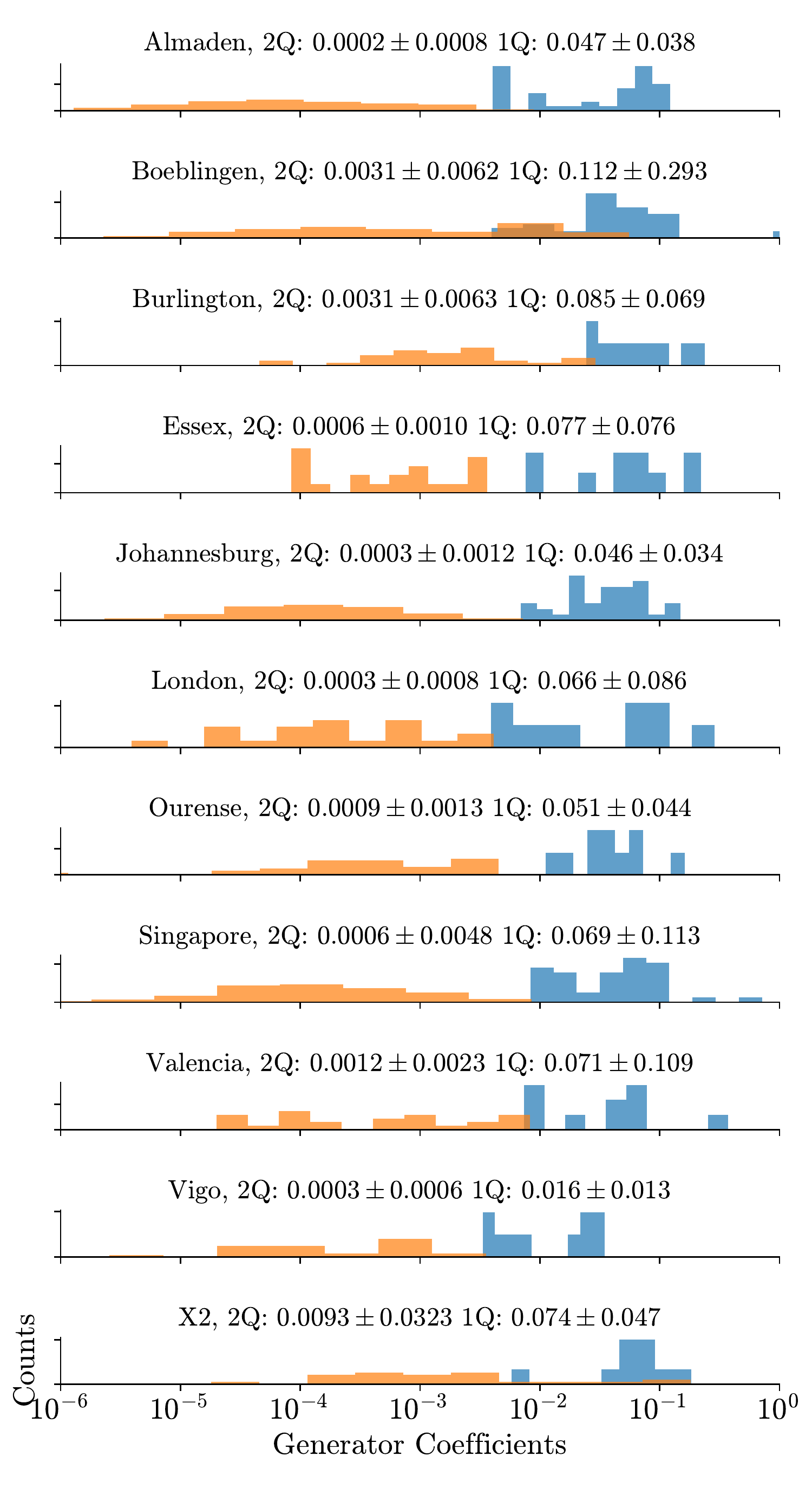}
    \caption{Generator coefficients for 1 (blue) and 2-qubit (orange) readout errors on several IBM  Quantum devices. In all cases, the rates are relatively consistent. Some error rates are calculated as $r_i=0$ due to limitations in the number of shots. These cases are elided for the purposes of plotting with logarithmic axes.}
    \label{fig:comp_char}
    \end{figure}

    A natural expectation is that the correlated errors between qubits is dependent on their connectivity, and that correlated errors will be largest on neighbouring qubits.
    However, our results indicate that the correlation in measurement errors between distant qubits is non-trivial, and in some cases comparable to neighboring qubits.
  
    As one may expect, the decay generators have the highest associated generator coefficient. This is to be expected as thermal relaxation to the ground state is a significant contribution to measurement errors. However, it is surprising that the distribution median does not decay appreciably with qubit distance and is non-negligible even for distantly connected qubits.
    The excitation and exchange generators on the other hand show reduction with qubit distance on average, however in some cases certain rates can be as large as those between neighboring qubits.

  \subsection{Comparing Devices with CTMP-EM Data}\label{sec:devcomp}

    To illustrate the usefulness of CTMP-EM for characterization, we use this method to characterize readout errors on several IBM Quantum devices and compare their local 1-qubit and correlated 2-qubit generator coefficients. The distribution in generator values is shown in \cref{fig:comp_char}.
    For all measured devices the 1-qubit error rates are generally higher than the 2-qubit error rates as expected, and all devices give relatively consistent error rates. The devices shown range from 5-qubit to 20-qubits with the calibrations run using the minimum number of $n+2$ calibration circuits, using the maximum number of shots available for each device.
    Moreover, CTMP-EM as a characterization technique does not depend on device connectivity, and includes information about long-range correlations. This is advantageous since we saw before that the long-range correlations in readout errors can be non-trivial.

\section{Conclusion}\label{sec:conclusion}
  In this paper, we have shown that CTMP-EM is vital for improving the performance of VQAs on near-term devices, and that it can fundamentally change the shape of the objective function computed by the quantum device. Moreover, we demonstrate that CTMP-EM can efficiently be used to characterize long-range correlations in readout error on near-term devices in terms of the device connectivity, and that these long-range correlations are present on current devices.
  Nevertheless, an interesting topic for future work would be to expand the CTMP model to generators that act on more than two qubits, or analyze how the calibration parameters drift over time. Additionally, it would be interesting to investigate the impact on performance for QAOA Hamiltonians.
  For now, our work emphasizes the importance of using CTMP-EM for characterization, because it only requires computing the terms of $G$, of which there are only $O(n^2)$ many, instead of $A = e^G$, which is dense. Moreover, the calibration technique is efficient in the number of circuits.
  We believe that our results will be useful to understanding the objective function in VQAs on near-term devices, as well as characterizing near-term devices in terms of readout error. 

\section*{Acknowledgements}
    The authors thank Sergey Bravyi, Sophia Economou, and Sarah Sheldon for helpful discussions. We also thank the IBM Quantum team for providing access to the devices used in this work. This work was done during G.S.B internship at IBM Quantum during the summer of 2020. G.S.B. thanks the IBM Quantum team for a very enriching internship experience. 

\bibliography{bib}

\begin{thebibliography}{41}%
\makeatletter
\providecommand \@ifxundefined [1]{%
 \@ifx{#1\undefined}
}%
\providecommand \@ifnum [1]{%
 \ifnum #1\expandafter \@firstoftwo
 \else \expandafter \@secondoftwo
 \fi
}%
\providecommand \@ifx [1]{%
 \ifx #1\expandafter \@firstoftwo
 \else \expandafter \@secondoftwo
 \fi
}%
\providecommand \natexlab [1]{#1}%
\providecommand \enquote  [1]{``#1''}%
\providecommand \bibnamefont  [1]{#1}%
\providecommand \bibfnamefont [1]{#1}%
\providecommand \citenamefont [1]{#1}%
\providecommand \href@noop [0]{\@secondoftwo}%
\providecommand \href [0]{\begingroup \@sanitize@url \@href}%
\providecommand \@href[1]{\@@startlink{#1}\@@href}%
\providecommand \@@href[1]{\endgroup#1\@@endlink}%
\providecommand \@sanitize@url [0]{\catcode `\\12\catcode `\$12\catcode
  `\&12\catcode `\#12\catcode `\^12\catcode `\_12\catcode `\%12\relax}%
\providecommand \@@startlink[1]{}%
\providecommand \@@endlink[0]{}%
\providecommand \url  [0]{\begingroup\@sanitize@url \@url }%
\providecommand \@url [1]{\endgroup\@href {#1}{\urlprefix }}%
\providecommand \urlprefix  [0]{URL }%
\providecommand \Eprint [0]{\href }%
\providecommand \doibase [0]{http://dx.doi.org/}%
\providecommand \selectlanguage [0]{\@gobble}%
\providecommand \bibinfo  [0]{\@secondoftwo}%
\providecommand \bibfield  [0]{\@secondoftwo}%
\providecommand \translation [1]{[#1]}%
\providecommand \BibitemOpen [0]{}%
\providecommand \bibitemStop [0]{}%
\providecommand \bibitemNoStop [0]{.\EOS\space}%
\providecommand \EOS [0]{\spacefactor3000\relax}%
\providecommand \BibitemShut  [1]{\csname bibitem#1\endcsname}%
\let\auto@bib@innerbib\@empty
\bibitem [{\citenamefont {Cross}\ \emph {et~al.}(2019)\citenamefont {Cross},
  \citenamefont {Bishop}, \citenamefont {Sheldon}, \citenamefont {Nation},\
  and\ \citenamefont {Gambetta}}]{cross2019validating}%
  \BibitemOpen
  \bibfield  {author} {\bibinfo {author} {\bibfnamefont {Andrew~W.}\
  \bibnamefont {Cross}}, \bibinfo {author} {\bibfnamefont {Lev~S.}\
  \bibnamefont {Bishop}}, \bibinfo {author} {\bibfnamefont {Sarah}\
  \bibnamefont {Sheldon}}, \bibinfo {author} {\bibfnamefont {Paul~D.}\
  \bibnamefont {Nation}}, \ and\ \bibinfo {author} {\bibfnamefont {Jay~M.}\
  \bibnamefont {Gambetta}},\ }\bibfield  {title} {\enquote {\bibinfo {title}
  {Validating quantum computers using randomized model circuits},}\ }\href
  {\doibase 10.1103/PhysRevA.100.032328} {\bibfield  {journal} {\bibinfo
  {journal} {Phys. Rev. A}\ }\textbf {\bibinfo {volume} {100}},\ \bibinfo
  {pages} {032328} (\bibinfo {year} {2019})}\BibitemShut {NoStop}%
\bibitem [{\citenamefont {Pino}\ \emph {et~al.}(2020)\citenamefont {Pino},
  \citenamefont {Dreiling}, \citenamefont {Figgatt}, \citenamefont {Gaebler},
  \citenamefont {Moses}, \citenamefont {Baldwin}, \citenamefont {Foss-Feig},
  \citenamefont {Hayes}, \citenamefont {Mayer}, \citenamefont {Ryan-Anderson},\
  and\ \citenamefont {et~al.}}]{pino2020demonstration}%
  \BibitemOpen
  \bibfield  {author} {\bibinfo {author} {\bibfnamefont {J.~M.}\ \bibnamefont
  {Pino}}, \bibinfo {author} {\bibfnamefont {J.~M.}\ \bibnamefont {Dreiling}},
  \bibinfo {author} {\bibfnamefont {C.}~\bibnamefont {Figgatt}}, \bibinfo
  {author} {\bibfnamefont {J.~P.}\ \bibnamefont {Gaebler}}, \bibinfo {author}
  {\bibfnamefont {S.~A.}\ \bibnamefont {Moses}}, \bibinfo {author}
  {\bibfnamefont {C.~H.}\ \bibnamefont {Baldwin}}, \bibinfo {author}
  {\bibfnamefont {M.}~\bibnamefont {Foss-Feig}}, \bibinfo {author}
  {\bibfnamefont {D.}~\bibnamefont {Hayes}}, \bibinfo {author} {\bibfnamefont
  {K.}~\bibnamefont {Mayer}}, \bibinfo {author} {\bibfnamefont
  {C.}~\bibnamefont {Ryan-Anderson}}, \ and\ \bibinfo {author} {\bibnamefont
  {et~al.}},\ }\bibfield  {title} {\enquote {\bibinfo {title} {Demonstration of
  the qccd trapped-ion quantum computer architecture},}\ }\href
  {https://arxiv.org/abs/2003.01293} {\bibfield  {journal} {\bibinfo  {journal}
  {arXiv e-prints}\ ,\ \bibinfo {pages} {2003.01293}} (\bibinfo {year}
  {2020})}\BibitemShut {NoStop}%
\bibitem [{\citenamefont {Karalekas}\ \emph {et~al.}(2020)\citenamefont
  {Karalekas}, \citenamefont {Tezak}, \citenamefont {Peterson}, \citenamefont
  {Ryan}, \citenamefont {da~Silva},\ and\ \citenamefont
  {Smith}}]{karalekas2020quantum}%
  \BibitemOpen
  \bibfield  {author} {\bibinfo {author} {\bibfnamefont {Peter~J}\ \bibnamefont
  {Karalekas}}, \bibinfo {author} {\bibfnamefont {Nikolas~A}\ \bibnamefont
  {Tezak}}, \bibinfo {author} {\bibfnamefont {Eric~C}\ \bibnamefont
  {Peterson}}, \bibinfo {author} {\bibfnamefont {Colm~A}\ \bibnamefont {Ryan}},
  \bibinfo {author} {\bibfnamefont {Marcus~P}\ \bibnamefont {da~Silva}}, \ and\
  \bibinfo {author} {\bibfnamefont {Robert~S}\ \bibnamefont {Smith}},\
  }\bibfield  {title} {\enquote {\bibinfo {title} {A quantum-classical cloud
  platform optimized for variational hybrid algorithms},}\ }\href {\doibase
  10.1088/2058-9565/ab7559} {\bibfield  {journal} {\bibinfo  {journal} {Quantum
  Science and Technology}\ }\textbf {\bibinfo {volume} {5}},\ \bibinfo {pages}
  {024003} (\bibinfo {year} {2020})}\BibitemShut {NoStop}%
\bibitem [{\citenamefont {Jurcevic}\ \emph {et~al.}(2020)\citenamefont
  {Jurcevic}, \citenamefont {Javadi-Abhari}, \citenamefont {Bishop},
  \citenamefont {Lauer}, \citenamefont {Bogorin}, \citenamefont {Brink},
  \citenamefont {Capelluto}, \citenamefont {Günlük}, \citenamefont {Itoko},
  \citenamefont {Kanazawa}, \citenamefont {Kandala}, \citenamefont {Keefe},
  \citenamefont {Kruslich}, \citenamefont {Landers}, \citenamefont
  {Lewandowski}, \citenamefont {McClure}, \citenamefont {Nannicini},
  \citenamefont {Narasgond}, \citenamefont {Nayfeh}, \citenamefont {Pritchett},
  \citenamefont {Rothwell}, \citenamefont {Srinivasan}, \citenamefont
  {Sundaresan}, \citenamefont {Wang}, \citenamefont {Wei}, \citenamefont
  {Wood}, \citenamefont {Yau}, \citenamefont {Zhang}, \citenamefont {Dial},
  \citenamefont {Chow},\ and\ \citenamefont
  {Gambetta}}]{jurcevic2020demonstration}%
  \BibitemOpen
  \bibfield  {author} {\bibinfo {author} {\bibfnamefont {Petar}\ \bibnamefont
  {Jurcevic}}, \bibinfo {author} {\bibfnamefont {Ali}\ \bibnamefont
  {Javadi-Abhari}}, \bibinfo {author} {\bibfnamefont {Lev~S.}\ \bibnamefont
  {Bishop}}, \bibinfo {author} {\bibfnamefont {Isaac}\ \bibnamefont {Lauer}},
  \bibinfo {author} {\bibfnamefont {Daniela~F.}\ \bibnamefont {Bogorin}},
  \bibinfo {author} {\bibfnamefont {Markus}\ \bibnamefont {Brink}}, \bibinfo
  {author} {\bibfnamefont {Lauren}\ \bibnamefont {Capelluto}}, \bibinfo
  {author} {\bibfnamefont {Oktay}\ \bibnamefont {Günlük}}, \bibinfo {author}
  {\bibfnamefont {Toshinaro}\ \bibnamefont {Itoko}}, \bibinfo {author}
  {\bibfnamefont {Naoki}\ \bibnamefont {Kanazawa}}, \bibinfo {author}
  {\bibfnamefont {Abhinav}\ \bibnamefont {Kandala}}, \bibinfo {author}
  {\bibfnamefont {George~A.}\ \bibnamefont {Keefe}}, \bibinfo {author}
  {\bibfnamefont {Kevin}\ \bibnamefont {Kruslich}}, \bibinfo {author}
  {\bibfnamefont {William}\ \bibnamefont {Landers}}, \bibinfo {author}
  {\bibfnamefont {Eric~P.}\ \bibnamefont {Lewandowski}}, \bibinfo {author}
  {\bibfnamefont {Douglas~T.}\ \bibnamefont {McClure}}, \bibinfo {author}
  {\bibfnamefont {Giacomo}\ \bibnamefont {Nannicini}}, \bibinfo {author}
  {\bibfnamefont {Adinath}\ \bibnamefont {Narasgond}}, \bibinfo {author}
  {\bibfnamefont {Hasan~M.}\ \bibnamefont {Nayfeh}}, \bibinfo {author}
  {\bibfnamefont {Emily}\ \bibnamefont {Pritchett}}, \bibinfo {author}
  {\bibfnamefont {Mary~Beth}\ \bibnamefont {Rothwell}}, \bibinfo {author}
  {\bibfnamefont {Srikanth}\ \bibnamefont {Srinivasan}}, \bibinfo {author}
  {\bibfnamefont {Neereja}\ \bibnamefont {Sundaresan}}, \bibinfo {author}
  {\bibfnamefont {Cindy}\ \bibnamefont {Wang}}, \bibinfo {author}
  {\bibfnamefont {Ken~X.}\ \bibnamefont {Wei}}, \bibinfo {author}
  {\bibfnamefont {Christopher~J.}\ \bibnamefont {Wood}}, \bibinfo {author}
  {\bibfnamefont {Jeng-Bang}\ \bibnamefont {Yau}}, \bibinfo {author}
  {\bibfnamefont {Eric~J.}\ \bibnamefont {Zhang}}, \bibinfo {author}
  {\bibfnamefont {Oliver~E.}\ \bibnamefont {Dial}}, \bibinfo {author}
  {\bibfnamefont {Jerry~M.}\ \bibnamefont {Chow}}, \ and\ \bibinfo {author}
  {\bibfnamefont {Jay~M.}\ \bibnamefont {Gambetta}},\ }\bibfield  {title}
  {\enquote {\bibinfo {title} {Demonstration of quantum volume 64 on a
  superconducting quantum computing system},}\ }\href
  {https://arxiv.org/abs/2008.08571} {\  (\bibinfo {year} {2020})},\ \Eprint
  {http://arxiv.org/abs/2008.08571} {arXiv:2008.08571 [quant-ph]} \BibitemShut
  {NoStop}%
\bibitem [{\citenamefont {Farhi}\ \emph {et~al.}(2014)\citenamefont {Farhi},
  \citenamefont {Goldstone},\ and\ \citenamefont {Gutmann}}]{farhi2014quantum}%
  \BibitemOpen
  \bibfield  {author} {\bibinfo {author} {\bibfnamefont {Edward}\ \bibnamefont
  {Farhi}}, \bibinfo {author} {\bibfnamefont {Jeffrey}\ \bibnamefont
  {Goldstone}}, \ and\ \bibinfo {author} {\bibfnamefont {Sam}\ \bibnamefont
  {Gutmann}},\ }\bibfield  {title} {\enquote {\bibinfo {title} {A quantum
  approximate optimization algorithm},}\ }\href@noop {} {\bibfield  {journal}
  {\bibinfo  {journal} {arXiv preprint arXiv:1411.4028}\ } (\bibinfo {year}
  {2014})}\BibitemShut {NoStop}%
\bibitem [{\citenamefont {Peruzzo}\ \emph {et~al.}(2014)\citenamefont
  {Peruzzo}, \citenamefont {J.McClean}, \citenamefont {Shadbolt}, \citenamefont
  {M.-H.Yung}, \citenamefont {Zhou}, \citenamefont {Love}, \citenamefont
  {Aspuru-Guzik},\ and\ \citenamefont {O’Brien}}]{Peruzzo2014}%
  \BibitemOpen
  \bibfield  {author} {\bibinfo {author} {\bibfnamefont {A.}~\bibnamefont
  {Peruzzo}}, \bibinfo {author} {\bibnamefont {J.McClean}}, \bibinfo {author}
  {\bibfnamefont {P.}~\bibnamefont {Shadbolt}}, \bibinfo {author} {\bibnamefont
  {M.-H.Yung}}, \bibinfo {author} {\bibfnamefont {X.-Q.}\ \bibnamefont {Zhou}},
  \bibinfo {author} {\bibfnamefont {P.J.}\ \bibnamefont {Love}}, \bibinfo
  {author} {\bibfnamefont {A.}~\bibnamefont {Aspuru-Guzik}}, \ and\ \bibinfo
  {author} {\bibfnamefont {J.~L.}\ \bibnamefont {O’Brien}},\ }\bibfield
  {title} {\enquote {\bibinfo {title} {A variational eigenvalue solver on a
  photonic quantum processor},}\ }\href@noop {} {\bibfield  {journal} {\bibinfo
   {journal} {Nature Commun.}\ }\textbf {\bibinfo {volume} {5}},\ \bibinfo
  {pages} {4213} (\bibinfo {year} {2014})}\BibitemShut {NoStop}%
\bibitem [{\citenamefont {O'Malley}\ \emph {et~al.}(2016)\citenamefont
  {O'Malley}, \citenamefont {Babbush}, \citenamefont {Kivlichan}, \citenamefont
  {Romero}, \citenamefont {McClean}, \citenamefont {Barends}, \citenamefont
  {Kelly}, \citenamefont {Roushan}, \citenamefont {Tranter}, \citenamefont
  {Ding}, \citenamefont {Campbell}, \citenamefont {Chen}, \citenamefont {Chen},
  \citenamefont {Chiaro}, \citenamefont {Dunsworth}, \citenamefont {Fowler},
  \citenamefont {Jeffrey}, \citenamefont {Lucero}, \citenamefont {Megrant},
  \citenamefont {Mutus}, \citenamefont {Neeley}, \citenamefont {Neill},
  \citenamefont {Quintana}, \citenamefont {Sank}, \citenamefont {Vainsencher},
  \citenamefont {Wenner}, \citenamefont {White}, \citenamefont {Coveney},
  \citenamefont {Love}, \citenamefont {Neven}, \citenamefont {Aspuru-Guzik},\
  and\ \citenamefont {Martinis}}]{OMalley2016}%
  \BibitemOpen
  \bibfield  {author} {\bibinfo {author} {\bibfnamefont {P.~J.~J.}\
  \bibnamefont {O'Malley}}, \bibinfo {author} {\bibfnamefont {R.}~\bibnamefont
  {Babbush}}, \bibinfo {author} {\bibfnamefont {I.~D.}\ \bibnamefont
  {Kivlichan}}, \bibinfo {author} {\bibfnamefont {J.}~\bibnamefont {Romero}},
  \bibinfo {author} {\bibfnamefont {J.~R.}\ \bibnamefont {McClean}}, \bibinfo
  {author} {\bibfnamefont {R.}~\bibnamefont {Barends}}, \bibinfo {author}
  {\bibfnamefont {J.}~\bibnamefont {Kelly}}, \bibinfo {author} {\bibfnamefont
  {P.}~\bibnamefont {Roushan}}, \bibinfo {author} {\bibfnamefont
  {A.}~\bibnamefont {Tranter}}, \bibinfo {author} {\bibfnamefont
  {N.}~\bibnamefont {Ding}}, \bibinfo {author} {\bibfnamefont {B.}~\bibnamefont
  {Campbell}}, \bibinfo {author} {\bibfnamefont {Y.}~\bibnamefont {Chen}},
  \bibinfo {author} {\bibfnamefont {Z.}~\bibnamefont {Chen}}, \bibinfo {author}
  {\bibfnamefont {B.}~\bibnamefont {Chiaro}}, \bibinfo {author} {\bibfnamefont
  {A.}~\bibnamefont {Dunsworth}}, \bibinfo {author} {\bibfnamefont {A.~G.}\
  \bibnamefont {Fowler}}, \bibinfo {author} {\bibfnamefont {E.}~\bibnamefont
  {Jeffrey}}, \bibinfo {author} {\bibfnamefont {E.}~\bibnamefont {Lucero}},
  \bibinfo {author} {\bibfnamefont {A.}~\bibnamefont {Megrant}}, \bibinfo
  {author} {\bibfnamefont {J.~Y.}\ \bibnamefont {Mutus}}, \bibinfo {author}
  {\bibfnamefont {M.}~\bibnamefont {Neeley}}, \bibinfo {author} {\bibfnamefont
  {C.}~\bibnamefont {Neill}}, \bibinfo {author} {\bibfnamefont
  {C.}~\bibnamefont {Quintana}}, \bibinfo {author} {\bibfnamefont
  {D.}~\bibnamefont {Sank}}, \bibinfo {author} {\bibfnamefont {A.}~\bibnamefont
  {Vainsencher}}, \bibinfo {author} {\bibfnamefont {J.}~\bibnamefont {Wenner}},
  \bibinfo {author} {\bibfnamefont {T.~C.}\ \bibnamefont {White}}, \bibinfo
  {author} {\bibfnamefont {P.~V.}\ \bibnamefont {Coveney}}, \bibinfo {author}
  {\bibfnamefont {P.~J.}\ \bibnamefont {Love}}, \bibinfo {author}
  {\bibfnamefont {H.}~\bibnamefont {Neven}}, \bibinfo {author} {\bibfnamefont
  {A.}~\bibnamefont {Aspuru-Guzik}}, \ and\ \bibinfo {author} {\bibfnamefont
  {J.~M.}\ \bibnamefont {Martinis}},\ }\bibfield  {title} {\enquote {\bibinfo
  {title} {Scalable quantum simulation of molecular energies},}\ }\href
  {\doibase 10.1103/PhysRevX.6.031007} {\bibfield  {journal} {\bibinfo
  {journal} {Phys. Rev. X}\ }\textbf {\bibinfo {volume} {6}},\ \bibinfo {pages}
  {031007} (\bibinfo {year} {2016})}\BibitemShut {NoStop}%
\bibitem [{\citenamefont {McClean}\ \emph {et~al.}(2016)\citenamefont
  {McClean}, \citenamefont {Romero}, \citenamefont {Babbush},\ and\
  \citenamefont {Aspuru-Guzik}}]{McClean2016}%
  \BibitemOpen
  \bibfield  {author} {\bibinfo {author} {\bibfnamefont {Jarrod~R}\
  \bibnamefont {McClean}}, \bibinfo {author} {\bibfnamefont {Jonathan}\
  \bibnamefont {Romero}}, \bibinfo {author} {\bibfnamefont {Ryan}\ \bibnamefont
  {Babbush}}, \ and\ \bibinfo {author} {\bibfnamefont {Al{\'{a}}n}\
  \bibnamefont {Aspuru-Guzik}},\ }\bibfield  {title} {\enquote {\bibinfo
  {title} {The theory of variational hybrid quantum-classical algorithms},}\
  }\href {\doibase 10.1088/1367-2630/18/2/023023} {\bibfield  {journal}
  {\bibinfo  {journal} {New J. Phys}\ }\textbf {\bibinfo {volume} {18}},\
  \bibinfo {pages} {023023} (\bibinfo {year} {2016})}\BibitemShut {NoStop}%
\bibitem [{\citenamefont {Kandala}\ \emph {et~al.}(2017)\citenamefont
  {Kandala}, \citenamefont {Mezzacapo}, \citenamefont {Temme}, \citenamefont
  {Takita}, \citenamefont {Brink}, \citenamefont {Chow},\ and\ \citenamefont
  {Gambetta}}]{kandala2017hardware}%
  \BibitemOpen
  \bibfield  {author} {\bibinfo {author} {\bibfnamefont {Abhinav}\ \bibnamefont
  {Kandala}}, \bibinfo {author} {\bibfnamefont {Antonio}\ \bibnamefont
  {Mezzacapo}}, \bibinfo {author} {\bibfnamefont {Kristan}\ \bibnamefont
  {Temme}}, \bibinfo {author} {\bibfnamefont {Maika}\ \bibnamefont {Takita}},
  \bibinfo {author} {\bibfnamefont {Markus}\ \bibnamefont {Brink}}, \bibinfo
  {author} {\bibfnamefont {Jerry~M}\ \bibnamefont {Chow}}, \ and\ \bibinfo
  {author} {\bibfnamefont {Jay~M}\ \bibnamefont {Gambetta}},\ }\bibfield
  {title} {\enquote {\bibinfo {title} {Hardware-efficient variational quantum
  eigensolver for small molecules and quantum magnets},}\ }\href@noop {}
  {\bibfield  {journal} {\bibinfo  {journal} {Nature}\ }\textbf {\bibinfo
  {volume} {549}},\ \bibinfo {pages} {242--246} (\bibinfo {year}
  {2017})}\BibitemShut {NoStop}%
\bibitem [{\citenamefont {Colless}\ \emph {et~al.}(2018)\citenamefont
  {Colless}, \citenamefont {Ramasesh}, \citenamefont {Dahlen}, \citenamefont
  {Blok}, \citenamefont {Kimchi-Schwartz}, \citenamefont {McClean},
  \citenamefont {Carter}, \citenamefont {de~Jong},\ and\ \citenamefont
  {Siddiqi}}]{Colless2018}%
  \BibitemOpen
  \bibfield  {author} {\bibinfo {author} {\bibfnamefont {J.~I.}\ \bibnamefont
  {Colless}}, \bibinfo {author} {\bibfnamefont {V.~V.}\ \bibnamefont
  {Ramasesh}}, \bibinfo {author} {\bibfnamefont {D.}~\bibnamefont {Dahlen}},
  \bibinfo {author} {\bibfnamefont {M.~S.}\ \bibnamefont {Blok}}, \bibinfo
  {author} {\bibfnamefont {M.~E.}\ \bibnamefont {Kimchi-Schwartz}}, \bibinfo
  {author} {\bibfnamefont {J.~R.}\ \bibnamefont {McClean}}, \bibinfo {author}
  {\bibfnamefont {J.}~\bibnamefont {Carter}}, \bibinfo {author} {\bibfnamefont
  {W.~A.}\ \bibnamefont {de~Jong}}, \ and\ \bibinfo {author} {\bibfnamefont
  {I.}~\bibnamefont {Siddiqi}},\ }\bibfield  {title} {\enquote {\bibinfo
  {title} {Computation of molecular spectra on a quantum processor with an
  error-resilient algorithm},}\ }\href {\doibase 10.1103/PhysRevX.8.011021}
  {\bibfield  {journal} {\bibinfo  {journal} {Phys. Rev. X}\ }\textbf {\bibinfo
  {volume} {8}},\ \bibinfo {pages} {011021} (\bibinfo {year}
  {2018})}\BibitemShut {NoStop}%
\bibitem [{\citenamefont {Pagano}\ \emph {et~al.}(2019)\citenamefont {Pagano},
  \citenamefont {Bapat}, \citenamefont {Becker}, \citenamefont {Collins},
  \citenamefont {De}, \citenamefont {Hess}, \citenamefont {Kaplan},
  \citenamefont {Kyprianidis}, \citenamefont {Tan}, \citenamefont {Baldwin}
  \emph {et~al.}}]{pagano2019quantum}%
  \BibitemOpen
  \bibfield  {author} {\bibinfo {author} {\bibfnamefont {G}~\bibnamefont
  {Pagano}}, \bibinfo {author} {\bibfnamefont {A}~\bibnamefont {Bapat}},
  \bibinfo {author} {\bibfnamefont {P}~\bibnamefont {Becker}}, \bibinfo
  {author} {\bibfnamefont {KS}~\bibnamefont {Collins}}, \bibinfo {author}
  {\bibfnamefont {A}~\bibnamefont {De}}, \bibinfo {author} {\bibfnamefont
  {PW}~\bibnamefont {Hess}}, \bibinfo {author} {\bibfnamefont {HB}~\bibnamefont
  {Kaplan}}, \bibinfo {author} {\bibfnamefont {A}~\bibnamefont {Kyprianidis}},
  \bibinfo {author} {\bibfnamefont {WL}~\bibnamefont {Tan}}, \bibinfo {author}
  {\bibfnamefont {C}~\bibnamefont {Baldwin}},  \emph {et~al.},\ }\bibfield
  {title} {\enquote {\bibinfo {title} {Quantum approximate optimization with a
  trapped-ion quantum simulator},}\ }\href@noop {} {\bibfield  {journal}
  {\bibinfo  {journal} {arXiv preprint arXiv:1906.02700}\ } (\bibinfo {year}
  {2019})}\BibitemShut {NoStop}%
\bibitem [{\citenamefont {Romero}\ and\ \citenamefont
  {Aspuru-Guzik}(2019)}]{romero2019variational}%
  \BibitemOpen
  \bibfield  {author} {\bibinfo {author} {\bibfnamefont {Jonathan}\
  \bibnamefont {Romero}}\ and\ \bibinfo {author} {\bibfnamefont {Alan}\
  \bibnamefont {Aspuru-Guzik}},\ }\bibfield  {title} {\enquote {\bibinfo
  {title} {Variational quantum generators: Generative adversarial quantum
  machine learning for continuous distributions},}\ }\href@noop {} {\bibfield
  {journal} {\bibinfo  {journal} {arXiv preprint arXiv:1901.00848}\ } (\bibinfo
  {year} {2019})}\BibitemShut {NoStop}%
\bibitem [{\citenamefont {Ollitrault}\ \emph {et~al.}(2019)\citenamefont
  {Ollitrault}, \citenamefont {Kandala}, \citenamefont {Chen}, \citenamefont
  {Barkoutsos}, \citenamefont {Mezzacapo}, \citenamefont {Pistoia},
  \citenamefont {Sheldon}, \citenamefont {Woerner}, \citenamefont {Gambetta},\
  and\ \citenamefont {Tavernelli}}]{ollitrault2019quantum}%
  \BibitemOpen
  \bibfield  {author} {\bibinfo {author} {\bibfnamefont {Pauline~J}\
  \bibnamefont {Ollitrault}}, \bibinfo {author} {\bibfnamefont {Abhinav}\
  \bibnamefont {Kandala}}, \bibinfo {author} {\bibfnamefont {Chun-Fu}\
  \bibnamefont {Chen}}, \bibinfo {author} {\bibfnamefont {Panagiotis~Kl}\
  \bibnamefont {Barkoutsos}}, \bibinfo {author} {\bibfnamefont {Antonio}\
  \bibnamefont {Mezzacapo}}, \bibinfo {author} {\bibfnamefont {Marco}\
  \bibnamefont {Pistoia}}, \bibinfo {author} {\bibfnamefont {Sarah}\
  \bibnamefont {Sheldon}}, \bibinfo {author} {\bibfnamefont {Stefan}\
  \bibnamefont {Woerner}}, \bibinfo {author} {\bibfnamefont {Jay}\ \bibnamefont
  {Gambetta}}, \ and\ \bibinfo {author} {\bibfnamefont {Ivano}\ \bibnamefont
  {Tavernelli}},\ }\bibfield  {title} {\enquote {\bibinfo {title} {Quantum
  equation of motion for computing molecular excitation energies on a noisy
  quantum processor},}\ }\href {https://arxiv.org/abs/1910.12890} {\bibfield
  {journal} {\bibinfo  {journal} {arXiv preprint arXiv:1910.12890}\ } (\bibinfo
  {year} {2019})},\ \Eprint {http://arxiv.org/abs/1910.12890} {arXiv:1910.12890
  [quant-ph]} \BibitemShut {NoStop}%
\bibitem [{\citenamefont {{Arute}}\ \emph {et~al.}(2020)\citenamefont
  {{Arute}}, \citenamefont {{Arya}}, \citenamefont {{Babbush}}, \citenamefont
  {{Bacon}}, \citenamefont {{Bardin}}, \citenamefont {{Barends}}, \citenamefont
  {{Boixo}}, \citenamefont {{Broughton}}, \citenamefont {{Buckley}},
  \citenamefont {{Buell}}, \citenamefont {{Burkett}}, \citenamefont
  {{Bushnell}}, \citenamefont {{Chen}}, \citenamefont {{Chen}}, \citenamefont
  {{Chiaro}}, \citenamefont {{Collins}}, \citenamefont {{Courtney}},
  \citenamefont {{Demura}}, \citenamefont {{Dunsworth}}, \citenamefont
  {{Eppens}}, \citenamefont {{Farhi}}, \citenamefont {{Fowler}}, \citenamefont
  {{Foxen}}, \citenamefont {{Gidney}}, \citenamefont {{Giustina}},
  \citenamefont {{Graff}}, \citenamefont {{Habegger}}, \citenamefont
  {{Harrigan}}, \citenamefont {{Ho}}, \citenamefont {{Hong}}, \citenamefont
  {{Huang}}, \citenamefont {{Huggins}}, \citenamefont {{Ioffe}}, \citenamefont
  {{Isakov}}, \citenamefont {{Jeffrey}}, \citenamefont {{Jiang}}, \citenamefont
  {{Jones}}, \citenamefont {{Kafri}}, \citenamefont {{Kechedzhi}},
  \citenamefont {{Kelly}}, \citenamefont {{Kim}}, \citenamefont {{Klimov}},
  \citenamefont {{Korotkov}}, \citenamefont {{Kostritsa}}, \citenamefont
  {{Landhuis}}, \citenamefont {{Laptev}}, \citenamefont {{Lindmark}},
  \citenamefont {{Lucero}}, \citenamefont {{Martin}}, \citenamefont
  {{Martinis}}, \citenamefont {{McClean}}, \citenamefont {{McEwen}},
  \citenamefont {{Megrant}}, \citenamefont {{Mi}}, \citenamefont {{Mohseni}},
  \citenamefont {{Mruczkiewicz}}, \citenamefont {{Mutus}}, \citenamefont
  {{Naaman}}, \citenamefont {{Neeley}}, \citenamefont {{Neill}}, \citenamefont
  {{Neven}}, \citenamefont {{Yuezhen Niu}}, \citenamefont {{O'Brien}},
  \citenamefont {{Ostby}}, \citenamefont {{Petukhov}}, \citenamefont
  {{Putterman}}, \citenamefont {{Quintana}}, \citenamefont {{Roushan}},
  \citenamefont {{Rubin}}, \citenamefont {{Sank}}, \citenamefont {{Satzinger}},
  \citenamefont {{Smelyanskiy}}, \citenamefont {{Strain}}, \citenamefont
  {{Sung}}, \citenamefont {{Szalay}}, \citenamefont {{Takeshita}},
  \citenamefont {{Vainsencher}}, \citenamefont {{White}}, \citenamefont
  {{Wiebe}}, \citenamefont {{Yao}}, \citenamefont {{Yeh}},\ and\ \citenamefont
  {{Zalcman}}}]{arute2020hartree}%
  \BibitemOpen
  \bibfield  {author} {\bibinfo {author} {\bibfnamefont {Frank}\ \bibnamefont
  {{Arute}}}, \bibinfo {author} {\bibfnamefont {Kunal}\ \bibnamefont {{Arya}}},
  \bibinfo {author} {\bibfnamefont {Ryan}\ \bibnamefont {{Babbush}}}, \bibinfo
  {author} {\bibfnamefont {Dave}\ \bibnamefont {{Bacon}}}, \bibinfo {author}
  {\bibfnamefont {Joseph~C.}\ \bibnamefont {{Bardin}}}, \bibinfo {author}
  {\bibfnamefont {Rami}\ \bibnamefont {{Barends}}}, \bibinfo {author}
  {\bibfnamefont {Sergio}\ \bibnamefont {{Boixo}}}, \bibinfo {author}
  {\bibfnamefont {Michael}\ \bibnamefont {{Broughton}}}, \bibinfo {author}
  {\bibfnamefont {Bob~B.}\ \bibnamefont {{Buckley}}}, \bibinfo {author}
  {\bibfnamefont {David~A.}\ \bibnamefont {{Buell}}}, \bibinfo {author}
  {\bibfnamefont {Brian}\ \bibnamefont {{Burkett}}}, \bibinfo {author}
  {\bibfnamefont {Nicholas}\ \bibnamefont {{Bushnell}}}, \bibinfo {author}
  {\bibfnamefont {Yu}~\bibnamefont {{Chen}}}, \bibinfo {author} {\bibfnamefont
  {Zijun}\ \bibnamefont {{Chen}}}, \bibinfo {author} {\bibfnamefont {Benjamin}\
  \bibnamefont {{Chiaro}}}, \bibinfo {author} {\bibfnamefont {Roberto}\
  \bibnamefont {{Collins}}}, \bibinfo {author} {\bibfnamefont {William}\
  \bibnamefont {{Courtney}}}, \bibinfo {author} {\bibfnamefont {Sean}\
  \bibnamefont {{Demura}}}, \bibinfo {author} {\bibfnamefont {Andrew}\
  \bibnamefont {{Dunsworth}}}, \bibinfo {author} {\bibfnamefont {Daniel}\
  \bibnamefont {{Eppens}}}, \bibinfo {author} {\bibfnamefont {Edward}\
  \bibnamefont {{Farhi}}}, \bibinfo {author} {\bibfnamefont {Austin}\
  \bibnamefont {{Fowler}}}, \bibinfo {author} {\bibfnamefont {Brooks}\
  \bibnamefont {{Foxen}}}, \bibinfo {author} {\bibfnamefont {Craig}\
  \bibnamefont {{Gidney}}}, \bibinfo {author} {\bibfnamefont {Marissa}\
  \bibnamefont {{Giustina}}}, \bibinfo {author} {\bibfnamefont {Rob}\
  \bibnamefont {{Graff}}}, \bibinfo {author} {\bibfnamefont {Steve}\
  \bibnamefont {{Habegger}}}, \bibinfo {author} {\bibfnamefont {Matthew~P.}\
  \bibnamefont {{Harrigan}}}, \bibinfo {author} {\bibfnamefont {Alan}\
  \bibnamefont {{Ho}}}, \bibinfo {author} {\bibfnamefont {Sabrina}\
  \bibnamefont {{Hong}}}, \bibinfo {author} {\bibfnamefont {Trent}\
  \bibnamefont {{Huang}}}, \bibinfo {author} {\bibfnamefont {William~J.}\
  \bibnamefont {{Huggins}}}, \bibinfo {author} {\bibfnamefont {Lev}\
  \bibnamefont {{Ioffe}}}, \bibinfo {author} {\bibfnamefont {Sergei~V.}\
  \bibnamefont {{Isakov}}}, \bibinfo {author} {\bibfnamefont {Evan}\
  \bibnamefont {{Jeffrey}}}, \bibinfo {author} {\bibfnamefont {Zhang}\
  \bibnamefont {{Jiang}}}, \bibinfo {author} {\bibfnamefont {Cody}\
  \bibnamefont {{Jones}}}, \bibinfo {author} {\bibfnamefont {Dvir}\
  \bibnamefont {{Kafri}}}, \bibinfo {author} {\bibfnamefont {Kostyantyn}\
  \bibnamefont {{Kechedzhi}}}, \bibinfo {author} {\bibfnamefont {Julian}\
  \bibnamefont {{Kelly}}}, \bibinfo {author} {\bibfnamefont {Seon}\
  \bibnamefont {{Kim}}}, \bibinfo {author} {\bibfnamefont {Paul~V.}\
  \bibnamefont {{Klimov}}}, \bibinfo {author} {\bibfnamefont {Alexander}\
  \bibnamefont {{Korotkov}}}, \bibinfo {author} {\bibfnamefont {Fedor}\
  \bibnamefont {{Kostritsa}}}, \bibinfo {author} {\bibfnamefont {David}\
  \bibnamefont {{Landhuis}}}, \bibinfo {author} {\bibfnamefont {Pavel}\
  \bibnamefont {{Laptev}}}, \bibinfo {author} {\bibfnamefont {Mike}\
  \bibnamefont {{Lindmark}}}, \bibinfo {author} {\bibfnamefont {Erik}\
  \bibnamefont {{Lucero}}}, \bibinfo {author} {\bibfnamefont {Orion}\
  \bibnamefont {{Martin}}}, \bibinfo {author} {\bibfnamefont {John~M.}\
  \bibnamefont {{Martinis}}}, \bibinfo {author} {\bibfnamefont {Jarrod~R.}\
  \bibnamefont {{McClean}}}, \bibinfo {author} {\bibfnamefont {Matt}\
  \bibnamefont {{McEwen}}}, \bibinfo {author} {\bibfnamefont {Anthony}\
  \bibnamefont {{Megrant}}}, \bibinfo {author} {\bibfnamefont {Xiao}\
  \bibnamefont {{Mi}}}, \bibinfo {author} {\bibfnamefont {Masoud}\ \bibnamefont
  {{Mohseni}}}, \bibinfo {author} {\bibfnamefont {Wojciech}\ \bibnamefont
  {{Mruczkiewicz}}}, \bibinfo {author} {\bibfnamefont {Josh}\ \bibnamefont
  {{Mutus}}}, \bibinfo {author} {\bibfnamefont {Ofer}\ \bibnamefont
  {{Naaman}}}, \bibinfo {author} {\bibfnamefont {Matthew}\ \bibnamefont
  {{Neeley}}}, \bibinfo {author} {\bibfnamefont {Charles}\ \bibnamefont
  {{Neill}}}, \bibinfo {author} {\bibfnamefont {Hartmut}\ \bibnamefont
  {{Neven}}}, \bibinfo {author} {\bibfnamefont {Murphy}\ \bibnamefont {{Yuezhen
  Niu}}}, \bibinfo {author} {\bibfnamefont {Thomas~E.}\ \bibnamefont
  {{O'Brien}}}, \bibinfo {author} {\bibfnamefont {Eric}\ \bibnamefont
  {{Ostby}}}, \bibinfo {author} {\bibfnamefont {Andre}\ \bibnamefont
  {{Petukhov}}}, \bibinfo {author} {\bibfnamefont {Harald}\ \bibnamefont
  {{Putterman}}}, \bibinfo {author} {\bibfnamefont {Chris}\ \bibnamefont
  {{Quintana}}}, \bibinfo {author} {\bibfnamefont {Pedram}\ \bibnamefont
  {{Roushan}}}, \bibinfo {author} {\bibfnamefont {Nicholas~C.}\ \bibnamefont
  {{Rubin}}}, \bibinfo {author} {\bibfnamefont {Daniel}\ \bibnamefont
  {{Sank}}}, \bibinfo {author} {\bibfnamefont {Kevin~J.}\ \bibnamefont
  {{Satzinger}}}, \bibinfo {author} {\bibfnamefont {Vadim}\ \bibnamefont
  {{Smelyanskiy}}}, \bibinfo {author} {\bibfnamefont {Doug}\ \bibnamefont
  {{Strain}}}, \bibinfo {author} {\bibfnamefont {Kevin~J.}\ \bibnamefont
  {{Sung}}}, \bibinfo {author} {\bibfnamefont {Marco}\ \bibnamefont
  {{Szalay}}}, \bibinfo {author} {\bibfnamefont {Tyler~Y.}\ \bibnamefont
  {{Takeshita}}}, \bibinfo {author} {\bibfnamefont {Amit}\ \bibnamefont
  {{Vainsencher}}}, \bibinfo {author} {\bibfnamefont {Theodore}\ \bibnamefont
  {{White}}}, \bibinfo {author} {\bibfnamefont {Nathan}\ \bibnamefont
  {{Wiebe}}}, \bibinfo {author} {\bibfnamefont {Z.~Jamie}\ \bibnamefont
  {{Yao}}}, \bibinfo {author} {\bibfnamefont {Ping}\ \bibnamefont {{Yeh}}}, \
  and\ \bibinfo {author} {\bibfnamefont {Adam}\ \bibnamefont {{Zalcman}}},\
  }\bibfield  {title} {\enquote {\bibinfo {title} {{Hartree-Fock on a
  superconducting qubit quantum computer}},}\ }\href@noop {} {\bibfield
  {journal} {\bibinfo  {journal} {arXiv e-prints}\ ,\ \bibinfo {eid}
  {arXiv:2004.04174}} (\bibinfo {year} {2020})},\ \Eprint
  {http://arxiv.org/abs/2004.04174} {arXiv:2004.04174 [quant-ph]} \BibitemShut
  {NoStop}%
\bibitem [{\citenamefont {{Otterbach}}\ \emph {et~al.}(2017)\citenamefont
  {{Otterbach}}, \citenamefont {{Manenti}}, \citenamefont {{Alidoust}},
  \citenamefont {{Bestwick}}, \citenamefont {{Block}}, \citenamefont {{Bloom}},
  \citenamefont {{Caldwell}}, \citenamefont {{Didier}}, \citenamefont
  {{Schuyler Fried}}, \citenamefont {{Hong}}, \citenamefont {{Karalekas}},
  \citenamefont {{Osborn}}, \citenamefont {{Papageorge}}, \citenamefont
  {{Peterson}}, \citenamefont {{Prawiroatmodjo}}, \citenamefont {{Rubin}},
  \citenamefont {{Ryan}}, \citenamefont {{Scarabelli}}, \citenamefont
  {{Scheer}}, \citenamefont {{Sete}}, \citenamefont {{Sivarajah}},
  \citenamefont {{Smith}}, \citenamefont {{Staley}}, \citenamefont {{Tezak}},
  \citenamefont {{Zeng}}, \citenamefont {{Hudson}}, \citenamefont {{Johnson}},
  \citenamefont {{Reagor}}, \citenamefont {{da Silva}},\ and\ \citenamefont
  {{Rigetti}}}]{otterbach2017unsupervised}%
  \BibitemOpen
  \bibfield  {author} {\bibinfo {author} {\bibfnamefont {J.~S.}\ \bibnamefont
  {{Otterbach}}}, \bibinfo {author} {\bibfnamefont {R.}~\bibnamefont
  {{Manenti}}}, \bibinfo {author} {\bibfnamefont {N.}~\bibnamefont
  {{Alidoust}}}, \bibinfo {author} {\bibfnamefont {A.}~\bibnamefont
  {{Bestwick}}}, \bibinfo {author} {\bibfnamefont {M.}~\bibnamefont {{Block}}},
  \bibinfo {author} {\bibfnamefont {B.}~\bibnamefont {{Bloom}}}, \bibinfo
  {author} {\bibfnamefont {S.}~\bibnamefont {{Caldwell}}}, \bibinfo {author}
  {\bibfnamefont {N.}~\bibnamefont {{Didier}}}, \bibinfo {author}
  {\bibfnamefont {E.}~\bibnamefont {{Schuyler Fried}}}, \bibinfo {author}
  {\bibfnamefont {S.}~\bibnamefont {{Hong}}}, \bibinfo {author} {\bibfnamefont
  {P.}~\bibnamefont {{Karalekas}}}, \bibinfo {author} {\bibfnamefont {C.~B.}\
  \bibnamefont {{Osborn}}}, \bibinfo {author} {\bibfnamefont {A.}~\bibnamefont
  {{Papageorge}}}, \bibinfo {author} {\bibfnamefont {E.~C.}\ \bibnamefont
  {{Peterson}}}, \bibinfo {author} {\bibfnamefont {G.}~\bibnamefont
  {{Prawiroatmodjo}}}, \bibinfo {author} {\bibfnamefont {N.}~\bibnamefont
  {{Rubin}}}, \bibinfo {author} {\bibfnamefont {Colm~A.}\ \bibnamefont
  {{Ryan}}}, \bibinfo {author} {\bibfnamefont {D.}~\bibnamefont
  {{Scarabelli}}}, \bibinfo {author} {\bibfnamefont {M.}~\bibnamefont
  {{Scheer}}}, \bibinfo {author} {\bibfnamefont {E.~A.}\ \bibnamefont
  {{Sete}}}, \bibinfo {author} {\bibfnamefont {P.}~\bibnamefont {{Sivarajah}}},
  \bibinfo {author} {\bibfnamefont {Robert~S.}\ \bibnamefont {{Smith}}},
  \bibinfo {author} {\bibfnamefont {A.}~\bibnamefont {{Staley}}}, \bibinfo
  {author} {\bibfnamefont {N.}~\bibnamefont {{Tezak}}}, \bibinfo {author}
  {\bibfnamefont {W.~J.}\ \bibnamefont {{Zeng}}}, \bibinfo {author}
  {\bibfnamefont {A.}~\bibnamefont {{Hudson}}}, \bibinfo {author}
  {\bibfnamefont {Blake~R.}\ \bibnamefont {{Johnson}}}, \bibinfo {author}
  {\bibfnamefont {M.}~\bibnamefont {{Reagor}}}, \bibinfo {author}
  {\bibfnamefont {M.~P.}\ \bibnamefont {{da Silva}}}, \ and\ \bibinfo {author}
  {\bibfnamefont {C.}~\bibnamefont {{Rigetti}}},\ }\bibfield  {title} {\enquote
  {\bibinfo {title} {{Unsupervised Machine Learning on a Hybrid Quantum
  Computer}},}\ }\href@noop {} {\bibfield  {journal} {\bibinfo  {journal}
  {arXiv e-prints}\ ,\ \bibinfo {eid} {arXiv:1712.05771}} (\bibinfo {year}
  {2017})},\ \Eprint {http://arxiv.org/abs/1712.05771} {arXiv:1712.05771
  [quant-ph]} \BibitemShut {NoStop}%
\bibitem [{\citenamefont {Hadfield}\ \emph {et~al.}(2019)\citenamefont
  {Hadfield}, \citenamefont {Wang}, \citenamefont {O’Gorman}, \citenamefont
  {Rieffel}, \citenamefont {Venturelli},\ and\ \citenamefont
  {Biswas}}]{hadfield2019from}%
  \BibitemOpen
  \bibfield  {author} {\bibinfo {author} {\bibfnamefont {Stuart}\ \bibnamefont
  {Hadfield}}, \bibinfo {author} {\bibfnamefont {Zhihui}\ \bibnamefont {Wang}},
  \bibinfo {author} {\bibfnamefont {Bryan}\ \bibnamefont {O’Gorman}},
  \bibinfo {author} {\bibfnamefont {Eleanor}\ \bibnamefont {Rieffel}}, \bibinfo
  {author} {\bibfnamefont {Davide}\ \bibnamefont {Venturelli}}, \ and\ \bibinfo
  {author} {\bibfnamefont {Rupak}\ \bibnamefont {Biswas}},\ }\bibfield  {title}
  {\enquote {\bibinfo {title} {From the quantum approximate optimization
  algorithm to a quantum alternating operator ansatz},}\ }\href {\doibase
  10.3390/a12020034} {\bibfield  {journal} {\bibinfo  {journal} {Algorithms}\
  }\textbf {\bibinfo {volume} {12}},\ \bibinfo {pages} {34} (\bibinfo {year}
  {2019})}\BibitemShut {NoStop}%
\bibitem [{\citenamefont {Babbush}\ \emph {et~al.}(2018)\citenamefont
  {Babbush}, \citenamefont {Wiebe}, \citenamefont {McClean}, \citenamefont
  {McClain}, \citenamefont {Neven},\ and\ \citenamefont
  {Chan}}]{babbush2018low}%
  \BibitemOpen
  \bibfield  {author} {\bibinfo {author} {\bibfnamefont {Ryan}\ \bibnamefont
  {Babbush}}, \bibinfo {author} {\bibfnamefont {Nathan}\ \bibnamefont {Wiebe}},
  \bibinfo {author} {\bibfnamefont {Jarrod}\ \bibnamefont {McClean}}, \bibinfo
  {author} {\bibfnamefont {James}\ \bibnamefont {McClain}}, \bibinfo {author}
  {\bibfnamefont {Hartmut}\ \bibnamefont {Neven}}, \ and\ \bibinfo {author}
  {\bibfnamefont {Garnet Kin-Lic}\ \bibnamefont {Chan}},\ }\bibfield  {title}
  {\enquote {\bibinfo {title} {Low-depth quantum simulation of materials},}\
  }\href@noop {} {\bibfield  {journal} {\bibinfo  {journal} {Physical Review
  X}\ }\textbf {\bibinfo {volume} {8}},\ \bibinfo {pages} {011044} (\bibinfo
  {year} {2018})}\BibitemShut {NoStop}%
\bibitem [{\citenamefont {Verteletskyi}\ \emph {et~al.}(2020)\citenamefont
  {Verteletskyi}, \citenamefont {Yen},\ and\ \citenamefont
  {Izmaylov}}]{verteletskyi2019measurement}%
  \BibitemOpen
  \bibfield  {author} {\bibinfo {author} {\bibfnamefont {Vladyslav}\
  \bibnamefont {Verteletskyi}}, \bibinfo {author} {\bibfnamefont {Tzu-Ching}\
  \bibnamefont {Yen}}, \ and\ \bibinfo {author} {\bibfnamefont {Artur~F.}\
  \bibnamefont {Izmaylov}},\ }\bibfield  {title} {\enquote {\bibinfo {title}
  {Measurement optimization in the variational quantum eigensolver using a
  minimum clique cover},}\ }\href {\doibase 10.1063/1.5141458} {\bibfield
  {journal} {\bibinfo  {journal} {The Journal of Chemical Physics}\ }\textbf
  {\bibinfo {volume} {152}},\ \bibinfo {pages} {124114} (\bibinfo {year}
  {2020})},\ \Eprint {http://arxiv.org/abs/https://doi.org/10.1063/1.5141458}
  {https://doi.org/10.1063/1.5141458} \BibitemShut {NoStop}%
\bibitem [{\citenamefont {Huggins}\ \emph {et~al.}(2019)\citenamefont
  {Huggins}, \citenamefont {McClean}, \citenamefont {Rubin}, \citenamefont
  {Jiang}, \citenamefont {Wiebe}, \citenamefont {Whaley},\ and\ \citenamefont
  {Babbush}}]{huggins2019Efficient}%
  \BibitemOpen
  \bibfield  {author} {\bibinfo {author} {\bibfnamefont {William~J.}\
  \bibnamefont {Huggins}}, \bibinfo {author} {\bibfnamefont {Jarrod}\
  \bibnamefont {McClean}}, \bibinfo {author} {\bibfnamefont {Nicholas}\
  \bibnamefont {Rubin}}, \bibinfo {author} {\bibfnamefont {Zhang}\ \bibnamefont
  {Jiang}}, \bibinfo {author} {\bibfnamefont {Nathan}\ \bibnamefont {Wiebe}},
  \bibinfo {author} {\bibfnamefont {K.~Birgitta}\ \bibnamefont {Whaley}}, \
  and\ \bibinfo {author} {\bibfnamefont {Ryan}\ \bibnamefont {Babbush}},\
  }\bibfield  {title} {\enquote {\bibinfo {title} {Efficient and {{Noise
  Resilient Measurements}} for {{Quantum Chemistry}} on {{Near}}-{{Term Quantum
  Computers}}},}\ }\href@noop {} {\bibfield  {journal} {\bibinfo  {journal}
  {arXiv:1907.13117 [physics, physics:quant-ph]}\ } (\bibinfo {year} {2019})},\
  \Eprint {http://arxiv.org/abs/1907.13117} {arXiv:1907.13117 [physics,
  physics:quant-ph]} \BibitemShut {NoStop}%
\bibitem [{\citenamefont {Zhao}\ \emph {et~al.}(2020)\citenamefont {Zhao},
  \citenamefont {Tranter}, \citenamefont {Kirby}, \citenamefont {Ung},
  \citenamefont {Miyake},\ and\ \citenamefont {Love}}]{zhao2020measurement}%
  \BibitemOpen
  \bibfield  {author} {\bibinfo {author} {\bibfnamefont {Andrew}\ \bibnamefont
  {Zhao}}, \bibinfo {author} {\bibfnamefont {Andrew}\ \bibnamefont {Tranter}},
  \bibinfo {author} {\bibfnamefont {William~M.}\ \bibnamefont {Kirby}},
  \bibinfo {author} {\bibfnamefont {Shu~Fay}\ \bibnamefont {Ung}}, \bibinfo
  {author} {\bibfnamefont {Akimasa}\ \bibnamefont {Miyake}}, \ and\ \bibinfo
  {author} {\bibfnamefont {Peter~J.}\ \bibnamefont {Love}},\ }\bibfield
  {title} {\enquote {\bibinfo {title} {Measurement reduction in variational
  quantum algorithms},}\ }\href {\doibase 10.1103/PhysRevA.101.062322}
  {\bibfield  {journal} {\bibinfo  {journal} {Phys. Rev. A}\ }\textbf {\bibinfo
  {volume} {101}},\ \bibinfo {pages} {062322} (\bibinfo {year}
  {2020})}\BibitemShut {NoStop}%
\bibitem [{\citenamefont {Grimsley}\ \emph {et~al.}(2019)\citenamefont
  {Grimsley}, \citenamefont {Economou}, \citenamefont {Barnes},\ and\
  \citenamefont {Mayhall}}]{Grimsley2018}%
  \BibitemOpen
  \bibfield  {author} {\bibinfo {author} {\bibfnamefont {Harper~R.}\
  \bibnamefont {Grimsley}}, \bibinfo {author} {\bibfnamefont {Sophia~E.}\
  \bibnamefont {Economou}}, \bibinfo {author} {\bibfnamefont {Edwin}\
  \bibnamefont {Barnes}}, \ and\ \bibinfo {author} {\bibfnamefont
  {Nicholas~J.}\ \bibnamefont {Mayhall}},\ }\bibfield  {title} {\enquote
  {\bibinfo {title} {An adaptive variational algorithm for exact molecular
  simulations on a quantum computer},}\ }\href {\doibase
  10.1038/s41467-019-10988-2} {\bibfield  {journal} {\bibinfo  {journal} {Nat.
  Commun.}\ }\textbf {\bibinfo {volume} {10}},\ \bibinfo {pages} {3007}
  (\bibinfo {year} {2019})}\BibitemShut {NoStop}%
\bibitem [{\citenamefont {Barkoutsos}\ \emph {et~al.}(2018)\citenamefont
  {Barkoutsos}, \citenamefont {Gonthier}, \citenamefont {Sokolov},
  \citenamefont {Moll}, \citenamefont {Salis}, \citenamefont {Fuhrer},
  \citenamefont {Ganzhorn}, \citenamefont {Egger}, \citenamefont {Troyer},
  \citenamefont {Mezzacapo}, \citenamefont {Filipp},\ and\ \citenamefont
  {Tavernelli}}]{Barkoutsos2018}%
  \BibitemOpen
  \bibfield  {author} {\bibinfo {author} {\bibfnamefont {Panagiotis~Kl.}\
  \bibnamefont {Barkoutsos}}, \bibinfo {author} {\bibfnamefont {Jerome~F.}\
  \bibnamefont {Gonthier}}, \bibinfo {author} {\bibfnamefont {Igor}\
  \bibnamefont {Sokolov}}, \bibinfo {author} {\bibfnamefont {Nikolaj}\
  \bibnamefont {Moll}}, \bibinfo {author} {\bibfnamefont {Gian}\ \bibnamefont
  {Salis}}, \bibinfo {author} {\bibfnamefont {Andreas}\ \bibnamefont {Fuhrer}},
  \bibinfo {author} {\bibfnamefont {Marc}\ \bibnamefont {Ganzhorn}}, \bibinfo
  {author} {\bibfnamefont {Daniel~J.}\ \bibnamefont {Egger}}, \bibinfo {author}
  {\bibfnamefont {Matthias}\ \bibnamefont {Troyer}}, \bibinfo {author}
  {\bibfnamefont {Antonio}\ \bibnamefont {Mezzacapo}}, \bibinfo {author}
  {\bibfnamefont {Stefan}\ \bibnamefont {Filipp}}, \ and\ \bibinfo {author}
  {\bibfnamefont {Ivano}\ \bibnamefont {Tavernelli}},\ }\bibfield  {title}
  {\enquote {\bibinfo {title} {Quantum algorithms for electronic structure
  calculations: Particle-hole hamiltonian and optimized wave-function
  expansions},}\ }\href {\doibase 10.1103/PhysRevA.98.022322} {\bibfield
  {journal} {\bibinfo  {journal} {Phys. Rev. A}\ }\textbf {\bibinfo {volume}
  {98}},\ \bibinfo {pages} {022322} (\bibinfo {year} {2018})}\BibitemShut
  {NoStop}%
\bibitem [{\citenamefont {Tang}\ \emph {et~al.}(2019)\citenamefont {Tang},
  \citenamefont {Shkolnikov}, \citenamefont {Barron}, \citenamefont {Grimsley},
  \citenamefont {Mayhall}, \citenamefont {Barnes},\ and\ \citenamefont
  {Economou}}]{tang2019qubitadaptvqe}%
  \BibitemOpen
  \bibfield  {author} {\bibinfo {author} {\bibfnamefont {Ho~Lun}\ \bibnamefont
  {Tang}}, \bibinfo {author} {\bibfnamefont {V.~O.}\ \bibnamefont
  {Shkolnikov}}, \bibinfo {author} {\bibfnamefont {George~S.}\ \bibnamefont
  {Barron}}, \bibinfo {author} {\bibfnamefont {Harper~R.}\ \bibnamefont
  {Grimsley}}, \bibinfo {author} {\bibfnamefont {Nicholas~J.}\ \bibnamefont
  {Mayhall}}, \bibinfo {author} {\bibfnamefont {Edwin}\ \bibnamefont {Barnes}},
  \ and\ \bibinfo {author} {\bibfnamefont {Sophia~E.}\ \bibnamefont
  {Economou}},\ }\href {https://arxiv.org/abs/1911.10205} {\enquote {\bibinfo
  {title} {qubit-adapt-vqe: An adaptive algorithm for constructing
  hardware-efficient ansatze on a quantum processor},}\ } (\bibinfo {year}
  {2019}),\ \Eprint {http://arxiv.org/abs/1911.10205} {arXiv:1911.10205
  [quant-ph]} \BibitemShut {NoStop}%
\bibitem [{\citenamefont {Gard}\ \emph {et~al.}(2019)\citenamefont {Gard},
  \citenamefont {Zhu}, \citenamefont {Barron}, \citenamefont {Mayhall},
  \citenamefont {Economou},\ and\ \citenamefont {Barnes}}]{gard2019efficient}%
  \BibitemOpen
  \bibfield  {author} {\bibinfo {author} {\bibfnamefont {Bryan~T}\ \bibnamefont
  {Gard}}, \bibinfo {author} {\bibfnamefont {Linghua}\ \bibnamefont {Zhu}},
  \bibinfo {author} {\bibfnamefont {George~S}\ \bibnamefont {Barron}}, \bibinfo
  {author} {\bibfnamefont {Nicholas~J}\ \bibnamefont {Mayhall}}, \bibinfo
  {author} {\bibfnamefont {Sophia~E}\ \bibnamefont {Economou}}, \ and\ \bibinfo
  {author} {\bibfnamefont {Edwin}\ \bibnamefont {Barnes}},\ }\bibfield  {title}
  {\enquote {\bibinfo {title} {Efficient symmetry-preserving state preparation
  circuits for the variational quantum eigensolver algorithm},}\ }\href@noop {}
  {\bibfield  {journal} {\bibinfo  {journal} {arXiv preprint arXiv:1904.10910}\
  } (\bibinfo {year} {2019})}\BibitemShut {NoStop}%
\bibitem [{\citenamefont {Zhu}\ \emph {et~al.}(2019)\citenamefont {Zhu},
  \citenamefont {Linke}, \citenamefont {Benedetti}, \citenamefont {Landsman},
  \citenamefont {Nguyen}, \citenamefont {Alderete}, \citenamefont
  {Perdomo-Ortiz}, \citenamefont {Korda}, \citenamefont {Garfoot},
  \citenamefont {Brecque}, \citenamefont {Egan}, \citenamefont {Perdomo},\ and\
  \citenamefont {Monroe}}]{Zhueaaw9918}%
  \BibitemOpen
  \bibfield  {author} {\bibinfo {author} {\bibfnamefont {D.}~\bibnamefont
  {Zhu}}, \bibinfo {author} {\bibfnamefont {N.~M.}\ \bibnamefont {Linke}},
  \bibinfo {author} {\bibfnamefont {M.}~\bibnamefont {Benedetti}}, \bibinfo
  {author} {\bibfnamefont {K.~A.}\ \bibnamefont {Landsman}}, \bibinfo {author}
  {\bibfnamefont {N.~H.}\ \bibnamefont {Nguyen}}, \bibinfo {author}
  {\bibfnamefont {C.~H.}\ \bibnamefont {Alderete}}, \bibinfo {author}
  {\bibfnamefont {A.}~\bibnamefont {Perdomo-Ortiz}}, \bibinfo {author}
  {\bibfnamefont {N.}~\bibnamefont {Korda}}, \bibinfo {author} {\bibfnamefont
  {A.}~\bibnamefont {Garfoot}}, \bibinfo {author} {\bibfnamefont
  {C.}~\bibnamefont {Brecque}}, \bibinfo {author} {\bibfnamefont
  {L.}~\bibnamefont {Egan}}, \bibinfo {author} {\bibfnamefont {O.}~\bibnamefont
  {Perdomo}}, \ and\ \bibinfo {author} {\bibfnamefont {C.}~\bibnamefont
  {Monroe}},\ }\bibfield  {title} {\enquote {\bibinfo {title} {Training of
  quantum circuits on a hybrid quantum computer},}\ }\href {\doibase
  10.1126/sciadv.aaw9918} {\bibfield  {journal} {\bibinfo  {journal} {Science
  Advances}\ }\textbf {\bibinfo {volume} {5}} (\bibinfo {year} {2019}),\
  10.1126/sciadv.aaw9918},\ \Eprint
  {http://arxiv.org/abs/https://advances.sciencemag.org/content/5/10/eaaw9918.full.pdf}
  {https://advances.sciencemag.org/content/5/10/eaaw9918.full.pdf} \BibitemShut
  {NoStop}%
\bibitem [{\citenamefont {Dumitrescu}\ \emph {et~al.}(2018)\citenamefont
  {Dumitrescu}, \citenamefont {McCaskey}, \citenamefont {Hagen}, \citenamefont
  {Jansen}, \citenamefont {Morris}, \citenamefont {Papenbrock}, \citenamefont
  {Pooser}, \citenamefont {Dean},\ and\ \citenamefont
  {Lougovski}}]{dumitrescu2018cloud}%
  \BibitemOpen
  \bibfield  {author} {\bibinfo {author} {\bibfnamefont {E.~F.}\ \bibnamefont
  {Dumitrescu}}, \bibinfo {author} {\bibfnamefont {A.~J.}\ \bibnamefont
  {McCaskey}}, \bibinfo {author} {\bibfnamefont {G.}~\bibnamefont {Hagen}},
  \bibinfo {author} {\bibfnamefont {G.~R.}\ \bibnamefont {Jansen}}, \bibinfo
  {author} {\bibfnamefont {T.~D.}\ \bibnamefont {Morris}}, \bibinfo {author}
  {\bibfnamefont {T.}~\bibnamefont {Papenbrock}}, \bibinfo {author}
  {\bibfnamefont {R.~C.}\ \bibnamefont {Pooser}}, \bibinfo {author}
  {\bibfnamefont {D.~J.}\ \bibnamefont {Dean}}, \ and\ \bibinfo {author}
  {\bibfnamefont {P.}~\bibnamefont {Lougovski}},\ }\bibfield  {title} {\enquote
  {\bibinfo {title} {Cloud quantum computing of an atomic nucleus},}\ }\href
  {\doibase 10.1103/dumitrescu2018cloud} {\bibfield  {journal} {\bibinfo
  {journal} {Phys. Rev. Lett.}\ }\textbf {\bibinfo {volume} {120}},\ \bibinfo
  {pages} {210501} (\bibinfo {year} {2018})}\BibitemShut {NoStop}%
\bibitem [{\citenamefont {Klco}\ \emph {et~al.}(2018)\citenamefont {Klco},
  \citenamefont {Dumitrescu}, \citenamefont {McCaskey}, \citenamefont {Morris},
  \citenamefont {Pooser}, \citenamefont {Sanz}, \citenamefont {Solano},
  \citenamefont {Lougovski},\ and\ \citenamefont {Savage}}]{klco2018quantum}%
  \BibitemOpen
  \bibfield  {author} {\bibinfo {author} {\bibfnamefont {N.}~\bibnamefont
  {Klco}}, \bibinfo {author} {\bibfnamefont {E.~F.}\ \bibnamefont
  {Dumitrescu}}, \bibinfo {author} {\bibfnamefont {A.~J.}\ \bibnamefont
  {McCaskey}}, \bibinfo {author} {\bibfnamefont {T.~D.}\ \bibnamefont
  {Morris}}, \bibinfo {author} {\bibfnamefont {R.~C.}\ \bibnamefont {Pooser}},
  \bibinfo {author} {\bibfnamefont {M.}~\bibnamefont {Sanz}}, \bibinfo {author}
  {\bibfnamefont {E.}~\bibnamefont {Solano}}, \bibinfo {author} {\bibfnamefont
  {P.}~\bibnamefont {Lougovski}}, \ and\ \bibinfo {author} {\bibfnamefont
  {M.~J.}\ \bibnamefont {Savage}},\ }\bibfield  {title} {\enquote {\bibinfo
  {title} {Quantum-classical computation of schwinger model dynamics using
  quantum computers},}\ }\href {\doibase 10.1103/klco2018quantum} {\bibfield
  {journal} {\bibinfo  {journal} {Phys. Rev. A}\ }\textbf {\bibinfo {volume}
  {98}},\ \bibinfo {pages} {032331} (\bibinfo {year} {2018})}\BibitemShut
  {NoStop}%
\bibitem [{\citenamefont {Chen}\ \emph {et~al.}(2019)\citenamefont {Chen},
  \citenamefont {Farahzad}, \citenamefont {Yoo},\ and\ \citenamefont
  {Wei}}]{chen2019detector}%
  \BibitemOpen
  \bibfield  {author} {\bibinfo {author} {\bibfnamefont {Yanzhu}\ \bibnamefont
  {Chen}}, \bibinfo {author} {\bibfnamefont {Maziar}\ \bibnamefont {Farahzad}},
  \bibinfo {author} {\bibfnamefont {Shinjae}\ \bibnamefont {Yoo}}, \ and\
  \bibinfo {author} {\bibfnamefont {Tzu-Chieh}\ \bibnamefont {Wei}},\
  }\bibfield  {title} {\enquote {\bibinfo {title} {Detector tomography on ibm
  quantum computers and mitigation of an imperfect measurement},}\ }\href
  {\doibase 10.1103/PhysRevA.100.052315} {\bibfield  {journal} {\bibinfo
  {journal} {Phys. Rev. A}\ }\textbf {\bibinfo {volume} {100}},\ \bibinfo
  {pages} {052315} (\bibinfo {year} {2019})}\BibitemShut {NoStop}%
\bibitem [{\citenamefont {Temme}\ \emph {et~al.}(2017)\citenamefont {Temme},
  \citenamefont {Bravyi},\ and\ \citenamefont {Gambetta}}]{temme2017error}%
  \BibitemOpen
  \bibfield  {author} {\bibinfo {author} {\bibfnamefont {Kristan}\ \bibnamefont
  {Temme}}, \bibinfo {author} {\bibfnamefont {Sergey}\ \bibnamefont {Bravyi}},
  \ and\ \bibinfo {author} {\bibfnamefont {Jay~M.}\ \bibnamefont {Gambetta}},\
  }\bibfield  {title} {\enquote {\bibinfo {title} {Error mitigation for
  short-depth quantum circuits},}\ }\href {\doibase
  10.1103/PhysRevLett.119.180509} {\bibfield  {journal} {\bibinfo  {journal}
  {Phys. Rev. Lett.}\ }\textbf {\bibinfo {volume} {119}},\ \bibinfo {pages}
  {180509} (\bibinfo {year} {2017})}\BibitemShut {NoStop}%
\bibitem [{\citenamefont {Kandala}\ \emph {et~al.}(2019)\citenamefont
  {Kandala}, \citenamefont {Temme}, \citenamefont {C{\'o}rcoles}, \citenamefont
  {Mezzacapo}, \citenamefont {Chow},\ and\ \citenamefont
  {Gambetta}}]{kandala2019error}%
  \BibitemOpen
  \bibfield  {author} {\bibinfo {author} {\bibfnamefont {Abhinav}\ \bibnamefont
  {Kandala}}, \bibinfo {author} {\bibfnamefont {Kristan}\ \bibnamefont
  {Temme}}, \bibinfo {author} {\bibfnamefont {Antonio~D}\ \bibnamefont
  {C{\'o}rcoles}}, \bibinfo {author} {\bibfnamefont {Antonio}\ \bibnamefont
  {Mezzacapo}}, \bibinfo {author} {\bibfnamefont {Jerry~M}\ \bibnamefont
  {Chow}}, \ and\ \bibinfo {author} {\bibfnamefont {Jay~M}\ \bibnamefont
  {Gambetta}},\ }\bibfield  {title} {\enquote {\bibinfo {title} {Error
  mitigation extends the computational reach of a noisy quantum processor},}\
  }\href@noop {} {\bibfield  {journal} {\bibinfo  {journal} {Nature}\ }\textbf
  {\bibinfo {volume} {567}},\ \bibinfo {pages} {491--495} (\bibinfo {year}
  {2019})}\BibitemShut {NoStop}%
\bibitem [{\citenamefont {Hamilton}\ and\ \citenamefont
  {Pooser}(2020)}]{hamilton2020error}%
  \BibitemOpen
  \bibfield  {author} {\bibinfo {author} {\bibfnamefont {Kathleen~E.}\
  \bibnamefont {Hamilton}}\ and\ \bibinfo {author} {\bibfnamefont {Raphael~C.}\
  \bibnamefont {Pooser}},\ }\bibfield  {title} {\enquote {\bibinfo {title}
  {Error-mitigated data-driven circuit learning on noisy quantum hardware},}\
  }\href {\doibase 10/gg89qt} {\bibfield  {journal} {\bibinfo  {journal}
  {Quantum Machine Intelligence}\ }\textbf {\bibinfo {volume} {2}},\ \bibinfo
  {pages} {1–15} (\bibinfo {year} {2020})}\BibitemShut {NoStop}%
\bibitem [{\citenamefont {Geller}\ and\ \citenamefont
  {Sun}(2020)}]{geller2020efficient}%
  \BibitemOpen
  \bibfield  {author} {\bibinfo {author} {\bibfnamefont {Michael~R.}\
  \bibnamefont {Geller}}\ and\ \bibinfo {author} {\bibfnamefont {Mingyu}\
  \bibnamefont {Sun}},\ }\bibfield  {title} {\enquote {\bibinfo {title}
  {Efficient correction of multiqubit measurement errors},}\ }\href
  {https://arxiv.org/abs/2001.09980} {\bibfield  {journal} {\bibinfo  {journal}
  {arXiv preprint arXiv:2001.09980}\ } (\bibinfo {year} {2020})},\ \Eprint
  {http://arxiv.org/abs/2001.09980} {arXiv:2001.09980 [quant-ph]} \BibitemShut
  {NoStop}%
\bibitem [{\citenamefont {{Hamilton}}\ \emph {et~al.}(2020)\citenamefont
  {{Hamilton}}, \citenamefont {{Kharazi}}, \citenamefont {{Morris}},
  \citenamefont {{McCaskey}}, \citenamefont {{Bennink}},\ and\ \citenamefont
  {{Pooser}}}]{hamilton2020scalable}%
  \BibitemOpen
  \bibfield  {author} {\bibinfo {author} {\bibfnamefont {Kathleen~E.}\
  \bibnamefont {{Hamilton}}}, \bibinfo {author} {\bibfnamefont {Tyler}\
  \bibnamefont {{Kharazi}}}, \bibinfo {author} {\bibfnamefont {Titus}\
  \bibnamefont {{Morris}}}, \bibinfo {author} {\bibfnamefont {Alexander~J.}\
  \bibnamefont {{McCaskey}}}, \bibinfo {author} {\bibfnamefont {Ryan~S.}\
  \bibnamefont {{Bennink}}}, \ and\ \bibinfo {author} {\bibfnamefont
  {Raphael~C.}\ \bibnamefont {{Pooser}}},\ }\bibfield  {title} {\enquote
  {\bibinfo {title} {{Scalable quantum processor noise characterization}},}\
  }\href {https://arxiv.org/abs/2006.01805} {\bibfield  {journal} {\bibinfo
  {journal} {arXiv preprint arXiv:2006.01805}\ } (\bibinfo {year} {2020})},\
  \Eprint {http://arxiv.org/abs/2006.01805} {arXiv:2006.01805 [quant-ph]}
  \BibitemShut {NoStop}%
\bibitem [{\citenamefont {Geller}(2020)}]{geller2020rigorous}%
  \BibitemOpen
  \bibfield  {author} {\bibinfo {author} {\bibfnamefont {Michael~R}\
  \bibnamefont {Geller}},\ }\bibfield  {title} {\enquote {\bibinfo {title}
  {Rigorous measurement error correction},}\ }\href {\doibase
  10.1088/2058-9565/ab9591} {\bibfield  {journal} {\bibinfo  {journal} {Quantum
  Science and Technology}\ }\textbf {\bibinfo {volume} {5}},\ \bibinfo {pages}
  {03LT01} (\bibinfo {year} {2020})}\BibitemShut {NoStop}%
\bibitem [{\citenamefont {https://github.com/qiskit/qiskit}(2019)}]{Qiskit}%
  \BibitemOpen
  \bibfield  {author} {\bibinfo {author} {\bibnamefont
  {https://github.com/qiskit/qiskit}},\ }\href {\doibase
  10.5281/zenodo.2562110} {\enquote {\bibinfo {title} {Qiskit: An open-source
  framework for quantum computing},}\ } (\bibinfo {year} {2019})\BibitemShut
  {NoStop}%
\bibitem [{\citenamefont {{Bravyi}}\ \emph {et~al.}(2020)\citenamefont
  {{Bravyi}}, \citenamefont {{Sheldon}}, \citenamefont {{Kandala}},
  \citenamefont {{Mckay}},\ and\ \citenamefont {{Gambetta}}}]{bravyi}%
  \BibitemOpen
  \bibfield  {author} {\bibinfo {author} {\bibfnamefont {Sergey}\ \bibnamefont
  {{Bravyi}}}, \bibinfo {author} {\bibfnamefont {Sarah}\ \bibnamefont
  {{Sheldon}}}, \bibinfo {author} {\bibfnamefont {Abhinav}\ \bibnamefont
  {{Kandala}}}, \bibinfo {author} {\bibfnamefont {David~C.}\ \bibnamefont
  {{Mckay}}}, \ and\ \bibinfo {author} {\bibfnamefont {Jay~M.}\ \bibnamefont
  {{Gambetta}}},\ }\bibfield  {title} {\enquote {\bibinfo {title} {Mitigating
  measurement errors in multi-qubit experiments},}\ }\href
  {https://arxiv.org/abs/2006.14044} {\bibfield  {journal} {\bibinfo  {journal}
  {arXiv preprint arXiv:2006.14044}\ } (\bibinfo {year} {2020})},\ \Eprint
  {http://arxiv.org/abs/2006.14044} {arXiv:2006.14044 [quant-ph]} \BibitemShut
  {NoStop}%
\bibitem [{\citenamefont {Schuld}\ \emph {et~al.}(2019)\citenamefont {Schuld},
  \citenamefont {Bergholm}, \citenamefont {Gogolin}, \citenamefont {Izaac},\
  and\ \citenamefont {Killoran}}]{PhysRevA.99.032331}%
  \BibitemOpen
  \bibfield  {author} {\bibinfo {author} {\bibfnamefont {Maria}\ \bibnamefont
  {Schuld}}, \bibinfo {author} {\bibfnamefont {Ville}\ \bibnamefont
  {Bergholm}}, \bibinfo {author} {\bibfnamefont {Christian}\ \bibnamefont
  {Gogolin}}, \bibinfo {author} {\bibfnamefont {Josh}\ \bibnamefont {Izaac}}, \
  and\ \bibinfo {author} {\bibfnamefont {Nathan}\ \bibnamefont {Killoran}},\
  }\bibfield  {title} {\enquote {\bibinfo {title} {Evaluating analytic
  gradients on quantum hardware},}\ }\href {\doibase
  10.1103/PhysRevA.99.032331} {\bibfield  {journal} {\bibinfo  {journal} {Phys.
  Rev. A}\ }\textbf {\bibinfo {volume} {99}},\ \bibinfo {pages} {032331}
  (\bibinfo {year} {2019})}\BibitemShut {NoStop}%
\bibitem [{\citenamefont {Sharma}\ \emph {et~al.}(2020)\citenamefont {Sharma},
  \citenamefont {Khatri}, \citenamefont {Cerezo},\ and\ \citenamefont
  {Coles}}]{sharma2019noise}%
  \BibitemOpen
  \bibfield  {author} {\bibinfo {author} {\bibfnamefont {Kunal}\ \bibnamefont
  {Sharma}}, \bibinfo {author} {\bibfnamefont {Sumeet}\ \bibnamefont {Khatri}},
  \bibinfo {author} {\bibfnamefont {M}~\bibnamefont {Cerezo}}, \ and\ \bibinfo
  {author} {\bibfnamefont {Patrick~J}\ \bibnamefont {Coles}},\ }\bibfield
  {title} {\enquote {\bibinfo {title} {Noise resilience of variational quantum
  compiling},}\ }\href {\doibase 10.1088/1367-2630/ab784c} {\bibfield
  {journal} {\bibinfo  {journal} {New Journal of Physics}\ }\textbf {\bibinfo
  {volume} {22}},\ \bibinfo {pages} {043006} (\bibinfo {year}
  {2020})}\BibitemShut {NoStop}%
\bibitem [{\citenamefont {Hubbard}\ and\ \citenamefont
  {Flowers}(1963)}]{hubbard1963electron}%
  \BibitemOpen
  \bibfield  {author} {\bibinfo {author} {\bibfnamefont {J.}~\bibnamefont
  {Hubbard}}\ and\ \bibinfo {author} {\bibfnamefont {Brian~Hilton}\
  \bibnamefont {Flowers}},\ }\bibfield  {title} {\enquote {\bibinfo {title}
  {Electron correlations in narrow energy bands},}\ }\href {\doibase
  10.1098/rspa.1963.0204} {\bibfield  {journal} {\bibinfo  {journal}
  {Proceedings of the Royal Society of London. Series A. Mathematical and
  Physical Sciences}\ }\textbf {\bibinfo {volume} {276}},\ \bibinfo {pages}
  {238--257} (\bibinfo {year} {1963})},\ \Eprint
  {http://arxiv.org/abs/https://royalsocietypublishing.org/doi/pdf/10.1098/rspa.1963.0204}
  {https://royalsocietypublishing.org/doi/pdf/10.1098/rspa.1963.0204}
  \BibitemShut {NoStop}%
\bibitem [{\citenamefont {Bravyi}\ and\ \citenamefont
  {Kitaev}(2002)}]{BRAVYI2002210}%
  \BibitemOpen
  \bibfield  {author} {\bibinfo {author} {\bibfnamefont {Sergey~B.}\
  \bibnamefont {Bravyi}}\ and\ \bibinfo {author} {\bibfnamefont {Alexei~Yu.}\
  \bibnamefont {Kitaev}},\ }\bibfield  {title} {\enquote {\bibinfo {title}
  {Fermionic quantum computation},}\ }\href {\doibase
  https://doi.org/10.1006/aphy.2002.6254} {\bibfield  {journal} {\bibinfo
  {journal} {Annals of Physics}\ }\textbf {\bibinfo {volume} {298}},\ \bibinfo
  {pages} {210 -- 226} (\bibinfo {year} {2002})}\BibitemShut {NoStop}%
\bibitem [{\citenamefont {Bialczak}\ \emph {et~al.}(2010)\citenamefont
  {Bialczak}, \citenamefont {Ansmann}, \citenamefont {Hofheinz}, \citenamefont
  {Lucero}, \citenamefont {Neeley}, \citenamefont {O’Connell}, \citenamefont
  {Sank}, \citenamefont {Wang}, \citenamefont {Wenner}, \citenamefont
  {Steffen},\ and\ \citenamefont {et~al.}}]{bialczak2010quantum}%
  \BibitemOpen
  \bibfield  {author} {\bibinfo {author} {\bibfnamefont {R.~C.}\ \bibnamefont
  {Bialczak}}, \bibinfo {author} {\bibfnamefont {M.}~\bibnamefont {Ansmann}},
  \bibinfo {author} {\bibfnamefont {M.}~\bibnamefont {Hofheinz}}, \bibinfo
  {author} {\bibfnamefont {E.}~\bibnamefont {Lucero}}, \bibinfo {author}
  {\bibfnamefont {M.}~\bibnamefont {Neeley}}, \bibinfo {author} {\bibfnamefont
  {A.~D.}\ \bibnamefont {O’Connell}}, \bibinfo {author} {\bibfnamefont
  {D.}~\bibnamefont {Sank}}, \bibinfo {author} {\bibfnamefont {H.}~\bibnamefont
  {Wang}}, \bibinfo {author} {\bibfnamefont {J.}~\bibnamefont {Wenner}},
  \bibinfo {author} {\bibfnamefont {M.}~\bibnamefont {Steffen}}, \ and\
  \bibinfo {author} {\bibnamefont {et~al.}},\ }\bibfield  {title} {\enquote
  {\bibinfo {title} {Quantum process tomography of a universal entangling gate
  implemented with josephson phase qubits},}\ }\href {\doibase 10/dzfbqx}
  {\bibfield  {journal} {\bibinfo  {journal} {Nature Physics}\ }\textbf
  {\bibinfo {volume} {6}},\ \bibinfo {pages} {409–413} (\bibinfo {year}
  {2010})}\BibitemShut {NoStop}%
\end{thebibliography}%

\end{document}